# Numerical Models for the Diffuse Ionized Gas in Galaxies

## II. Three-dimensional radiative transfer in inhomogeneous interstellar structures as a tool for analyzing the diffuse ionized gas

J. A. Weber, A. W. A. Pauldrach, and T. L. Hoffmann

Institut für Astronomie und Astrophysik der Ludwig-Maximilians-Universität München, Scheinerstraße 1, 81679 München, Germany
{jweber, uh10107, hoffmann}@usm.lmu.de

October 23, 2018

**ABSTRACT**

*Context.* The diffuse ionized gas (DIG) constitutes the largest fraction of the total ionized interstellar matter in star-forming galaxies, but it is still unclear whether the ionization is driven predominantly by the ionizing radiation of hot massive stars, as in H II regions, or whether additional sources of ionization have to be considered. Key to understanding the ionization mechanisms in the DIG is the line emission by the ionized gas.
*Aims.* We systematically explore a plausible subset of the parameter space involving effective temperatures and metallicities of the ionizing sources, the effects of the hardening of their radiation by surrounding "leaky" H II regions with different escape fractions, as well as different scenarios for the clumpiness of the DIG, and compute the resulting line strength ratios for a number of diagnostic optical emission lines.
*Methods.* For the ionizing fluxes we compute a grid of stellar spectral energy distributions (SEDs) from detailed, fully non-LTE model atmospheres that include the effects of stellar winds and line blocking and blanketing. To calculate the ionization and temperature structure in the interstellar gas we use spherically symmetric photoionization models as well as state-of-the-art three-dimensional (3D) non-LTE radiative transfer simulations, considering hydrogen, helium, and the most abundant metals. We first apply these methods to classical H II regions around hot stars, using the model SEDs at different metallicities and effective temperatures as ionizing sources, and compute the spectral energy distributions of the escaping radiation for different escape fractions of hydrogen-ionizing photons. In a second step, we study the effects of the escaping radiation on the more dilute and extended DIG. Using 3D models simulating a section of a galactic spiral arm, we compute the ionization structure in the DIG for different scenarios for the inhomogeneity of the gas, assuming ionization by a stellar population SED based on plausible parameters.
*Results.* We provide quantitative predictions of how the line ratios from H II regions and the DIG vary as a function of metallicity $Z$, stellar effective temperature $T_{\rm eff}$, and escape fraction $f_{\rm esc}$ from the H II region. The range of predicted line ratios reinforces the hypothesis that the DIG is ionized by (filtered) radiation from hot stars. At one-tenth solar metallicity, radiation hardening is mostly due to hydrogen and helium, whereas at solar metallicity absorption by metals plays a significant role. The effects of hardening are seen primarily in the increase in the emission line ratios of the most important cooling lines of the gas, [N II]/H$\beta$ and [O II]/H$\beta$ at lower $T_{\rm eff}$, and [O III]/H$\beta$ at higher $T_{\rm eff}$. For low $T_{\rm eff}$ nearly the entire He I-ionizing radiation is absorbed in the H II regions, thereby preventing the formation of high ionization stages like O III in the DIG. The ionization structure of the DIG depends strongly on both the clumping factor $f_{\rm cl} = \langle n_{\rm H}^2 \rangle / \langle n_{\rm H} \rangle^2$ and the large-scale distribution of the gas. In our simulations about 10% of the ionizing radiation produced by hot massive stars in a spiral arm is sufficient to ionize the DIG up to a height of approximately 1 kpc above the galactic plane for a clumping factor close to the observed value of $f_{\rm cl} \sim 5$. Even small changes in simulation parameters like the clumping factor can lead to considerable variation in the ionized volume. Both for a more homogeneous gas and a very inhomogeneous gas containing both dense clumps and channels with low gas density, the ionized region in the dilute gas above the galactic plane can cease to be radiation-bounded, allowing the ionizing radiation to leak into the intergalactic medium. Comparison of observed and predicted line ratios indicates that the DIG is typically ionized with a softer SED than predicted by the chosen stellar population synthesis model.

## 1. Introduction

Besides cool molecular or atomic clouds with temperatures of approximately ten to a few hundred K, the warm neutral gas, and the very hot and thin fraction of the interstellar gas with temperatures on the order of one million K, which is heated, for instance, by supernova shocks, the interstellar medium contains a component of ionized material with an intermediate temperature of $T \approx 10\,000$ K. One part of this component are H II regions around hot stars (dwarf or supergiant stars of spectral class O or B, or central stars of planetary nebulae), but a significant amount of the mass and the line emission by intermediate-temperature gas is additionally contributed by dilute gas with a mean density of $\lesssim 0.1$ hydrogen ions per cubic centimeter, called *warm interstellar medium* (*WIM*, often used when referring to gas in the Milky Way) or *diffuse ionized gas* (*DIG*, usually referring to the diffuse gas in other galaxies), which can be found – in contrast to H II regions immediately connected to star formation – at heights of up to a few kpc above the galactic plane.[1]

---
[1] We will hereafter use the general term "DIG" to refer to the diffuse gas, and the expression "H II region" to describe the relatively dense ionized gas in the direct environment of hot massive stars. We note, however, that there is no well-defined distinction between those phases.

J. A. Weber et al.: 3D radiative transfer of inhomogeneous structuresThe nature of the energy sources that keep the DIG ionized is still not well understood. Photoionization by the radiation of the embedded sources is the dominant type of energy input for classical H II regions around hot stars; observations suggest that it also plays an important role for the ionization of the DIG.

For instance, Haffner et al. (1999) and Zurita et al. (2002) found a correlation between spiral arms and the presence of diffuse ionized gas. The ratio of the hydrogen recombination radiation photons originating from H II regions and the DIG has been determined observationally by Zurita et al. (2000), who found, based on an analysis of the observed emission measures in six nearby galaxies[2], that the total emission from the DIG is in the range of approximately 30% to 70% of the emission originating from the denser H II regions. Assuming ionization of the DIG predominantly by photons leaking from H II regions, this would require about half of the ionizing photons to escape from these H II regions.

Some of the spectral features found in the DIG are, however, considerably different from the features of dense H II regions ionized by embedded hot stars. Observations of the warm interstellar medium in our galaxy by Reynolds (1985) showed that the [O III]/H$\alpha$ line ratio is considerably smaller in the Galactic WIM than in typical H II regions. Reynolds & Tufte (1995) found only weak emission from the helium recombination line at 5876 Å, which indicates that the helium within the WIM of the Milky Way is mostly neutral. Measurements of the temperature of the gas, like the comparison of the nebular and auroral lines of N II performed by Reynolds et al. (2001), show temperatures clearly above the value characteristic for the gas in H II regions, which also leads to enhanced emission in the collisionally excited optical lines of singly ionized nitrogen and sulfur. An extensive study of the properties of the WIM in the solar environment of the Milky Way, based of observations using the Wisconsin H$\alpha$ Mapper spectrograph, has been presented by Madsen et al. (2006).

Observations by Greenawalt et al. (1997) find indications of a softer ionizing spectrum than in typical H II regions also for the DIG of M 31, although the difference appears to be less pronounced than in the Milky Way, as the spectra show [O III] emission as well as the presence of the helium recombination line at 5876 Å, which indicates an appreciable fraction of ionized helium. The dominance of only singly ionized metals in the DIG of the solar environment in the Milky Way cannot, however, be considered a general property of the entire DIG in spiral galaxies. Observations of edge-on galaxies (Otte et al. 2001, Collins & Rand 2001) show that in a part of these objects the [O III]/H$\alpha$ ratio rises at large heights above the galactic planes, while large [N II]/H$\alpha$ and [S II]/H$\alpha$ ratios are also observed. Evidence for a hardening of the ionizing spectra at high galactic altitudes has also been provided by Spitzer Space Telescope observations of three edge-on galaxies in the mid infrared by Rand et al. (2011), where a rising Ne III/Ne II ratio for larger distances from the galactic disks was found. The differences between the observations indicate that the energy sources, the temperature structure, and the ionization structure of the DIG may differ considerably from galaxy to galaxy or among different parts of the same galaxy.

Two main explanations for the differences between the spectra of classical H II regions and the DIG have been put forward, which are likely to be complementary: filtering effects modifying the spectral energy distribution of the ionizing radiation, and the presence of additional sources of ionization and heating. Filtering effects have been studied, for example, by Hoopes & Walterbos (2003) and Wood & Mathis (2004) and are partly able to explain the larger temperatures of the DIG (compared to H II regions) due to the radiation hardening effect. We will discuss filtering effects in more detail in Sect. 2.3.

An additional contribution of hot low-mass evolved stars (central stars of planetary nebulae or white dwarfs) has been discussed, for example, by Sokolowski & Bland-Hawthorn (1991) and Flores-Fajardo et al. (2011) to explain the fact that in some galaxies the [O III]/[O II] ratio increases for larger distances from the disk. Such objects can occur at larger distances from the plane of the star-forming galactic disk due to their older ages compared to hot massive main-sequence or supergiant stars.

A possible source for additional heating of the DIG is the dissipation of a part of the (macroscopic) mechanical energy of the gas. There have been several approaches for the explanation of the underlying mechanisms. One of them is shocks that can, for example, be produced by supernova explosions or strong stellar winds from massive OB associations (Collins & Rand 2001). Another possibility for providing the DIG with energy is via magnetic reconnection processes as described by Hoffmann et al. (2012), who have analyzed the contributions of photoionization and magnetic reconnection by computing synthetic spectra for purely photoionized or purely thermally (by magnetic reconnection) excited gas as well as for gas were both mechanisms are present. They find that while the total energy input provided by magnetic reconnection alone may be sufficient to keep the DIG ionized at large heights above the galactic plane, the observed spectra can be explained best by a mixture of both photoionization and thermal excitation.

Here, however, we will concentrate on the ability of hot massive stars close to the mid-plane of the galactic disk to ionize the DIG. Not much is known about the ionizing fluxes of these objects through direct observation. For this reason we employ realistic SEDs from state-of-the-art non-LTE atmospheric models that include the effects of stellar winds and line blocking and blanketing. The SEDs from our group have had a consistently good track record as ionizing sources in models of H II regions (Sellmaier et al. 1996; Giveon et al. 2002; Rubin et al. 2007, 2008, 2016). The use of such SEDs becomes even more important as we model gas and stars at different metallicities, where even fewer detailed direct and indirect observational constraints can be obtained from our nearer environment.

In this paper we systematically explore a plausible subset of the (rather large) parameter space in metallicity, stellar effective temperature, and filtering through surrounding H II regions, to show how observable characteristics of the DIG vary quantitatively with these parameters. As a parameter study, none of the models in our current paper is intended as a detailed description of the DIG in any particular galaxy, and neither is it expected that each particular parameter combination will be realized by nature somewhere in our immediate surroundings. We expect, however, that the large set of theoretical predictions that we present here will serve as a future guide in exploring more detailed quantitative models.

In Sect. 2 we study the properties of the DIG as a function of both the temperature- and metallicity-dependent SEDs of the ionizing sources – based on the model stars presented by Weber et al. (2015) – and the escape fractions from the H II regions surrounding the ionizing stars, using the spherically symmetric method described by Hoffmann et al. (2012) and Weber et al.

---

[2] The emission measure is defined as the integral of the square of the electron density along the line of sight, as most processes that determine the emissivity of the gas (recombination, collisional excitation, and thermal free-free emission) scale, for otherwise identical conditions, with the square of the gas density.

page 2 of 21

(2015). Together, our study comprises emission line ratios for a grid of 60 spherically symmetric DIG models.

A spherically symmetric approach is suitable for systematically examining the differential variations of the emission line spectrum in dependence of a certain input parameter, such as the spectral energy distribution of the ionizing source. There are, however, two drawbacks regarding simulations of the diffuse ionized gas. First, star-forming galaxies usually contain numerous H II regions whose stars may act as sources of ionization for more distant gas. Second, the density distribution of the absorbing gas is usually not spherically symmetric, and thus the attenuation of the radiation field is anisotropic. In Sect. 3 we present a set of 3D simulations whose geometry corresponds to a galactic region above a star-forming spiral arm, in order to account for these effects of the three-dimensional structure of the interstellar gas. Here we show the first synthetic emission line ratio maps of the DIG around a spiral arm with a spectral energy distribution based on a realistic IMF and compare the results with those obtained when assuming a softer ionizing spectrum corresponding to the SED of a model star with an effective temperature of 35 000 K. We also show how, even for the same mean density and average filling factor, the ionization structure of the DIG is affected by its porosity, which is a function of the size and the spatial distribution of the overdense "clumps".

## 2. Spherically symmetric simulations of dilute ionized gas surrounding leaky H II regions

In this section we describe a grid of H II region and DIG models in which we vary the effective temperatures $T_{\rm eff}$ and the metallicities $Z$ of the ionizing stars, as well the escape fraction of ionizing photons from the H II regions surrounding the stars into the DIG.[3] Our approach corresponds to that of Hoopes & Walterbos (2003), but our study is based on stellar SEDs from non-LTE atmospheric models taking into account the effects of stellar winds and line-blocking in the EUV, namely the model grid of Weber et al. (2015) at 0.1, 0.4, and 1.0 solar metallicities. We will first describe our numerical approach to accurately model the atmospheres of O-type stars that act as sources of ionization. Such detailed modelling is crucial for the determination of the radiation field in the ionizing spectral range above 1 Ryd, which in turn determines the temperature and ionization structure of the gas affected by the stellar radiation. Then we outline the method used to describe the processes in the ionized nebulae. After an explanation of the simulation setup used, we will discuss the effects of the filtering on the ionizing radiation field by the gas in the stellar environment and its effect on the emission line spectrum of both the H II regions in the stellar environment and the diffuse ionized gas. Finally we will compare our results with observations and discuss the implications of the comparison.

### 2.1. The general concept for calculating synthetic spectra and SEDs of massive stars

Our approach to modeling the expanding atmospheres of hot, massive stars has been described in detail in a series of previous papers (Pauldrach 1987, Pauldrach et al. 1990, 1993; Pauldrach et al. 1994, Taresch et al. 1997, Haser et al. 1998, Pauldrach et al. 1998, 2001, 2004, 2012), and we summarize the salient points here. Our method is based on the concept of homogeneous, stationary, and spherically symmetric radiation-driven atmospheres. Although this is an approximation to some extent, it is sufficient to reproduce all important characteristics of the expanding atmospheres in some detail.

A complete model atmosphere calculation consists of (*a*) a solution of the hydrodynamics describing velocity and density structures of the outflow, based on radiative acceleration by line, continuum, and Thomson absorption and scattering – an essential aspect of the model, because the expansion of the atmosphere alters the emergent flux considerably compared to a hydrostatic atmosphere; (*b*) the computation of the occupation numbers from a solution of the rate equations containing all important radiative and collisional processes, using sophisticated model atoms and corresponding line lists[4]; (*c*) a calculation of the radiation field from a detailed radiative transfer solution taking into account the Doppler-shifted line opacities and emissivities along with the continuum radiative transfer[5]; and (*d*) a computation of the temperature from the requirement of radiative (absorption/emission) and thermal (heating/cooling) balance. An accelerated Lambda iteration procedure[6] (ALImI, explained in detail by Pauldrach et al. 2014) is used to achieve consistency of occupation numbers, radiative transfer, and temperature. If required, an updated radiative acceleration can be computed from the converged model, and the process repeated.

---

[3] Unless stated otherwise, our use of the term "escape fraction" refers only to photons with energies above the hydrogen ionization threshold.

[4] In total 149 ionization stages of the 26 most abundant elements (H to Zn, apart from Li, Be, B, and Sc) are considered; a detailed description of the atomic models used is given in Sect. 3 and Table 1 of Pauldrach et al. 2001, and in Sect. 2 of Pauldrach et al. 1994 where several tables and figures that explain the overall procedure are shown. Low-temperature dielectronic recombination is included.

[5] If different spectral lines get shifted across the same observer's frame frequency by the velocity field in the envelope, line overlap, which is responsible for multiple-scattering events, takes place. The method used to solve this problem is an integral formulation of the transfer equation using an adaptive stepping technique on every ray (in $p, z$ geometry) in which the radiation transfer in each micro-interval is treated as a weighted sum on the microgrid,

$$I(\tau_0(p,z)) = I(\tau_n)e^{-(\tau_n-\tau_0)} + \sum_{i=0}^{n-1}\left(e^{-(\tau_i-\tau_0)}\int_{\tau_i}^{\tau_{i+1}} S(\tau)e^{-(\tau-\tau_i)}\,{\rm d}\tau(p,z)\right),$$

where $I$ is the specific intensity, $S$ is the source function and $\tau$ is the optical depth. To accurately account for the variation of the line opacities and emissivities due to the Doppler shift, all line profile functions are evaluated correctly for the current microgrid-coordinates on the ray, thus effectively resolving individual line profiles (cf. Pauldrach et al. 2001); thus, the effects of line overlap and multiple scattering are naturally included. On the basis of this procedure, the application of the Sobolev technique yields for the radiative line acceleration

$$g_{\rm lines}(r) = \frac{2\pi}{c}\frac{1}{\rho(r)}\sum_{\rm lines}\chi_{\rm line}(r)\int_{-1}^{+1} I_{\nu_0}(r,\mu)\frac{1-e^{-\tau_s(r,\mu)}}{\tau_s(r,\mu)}\mu\,{\rm d}\mu,$$

where

$$\tau_s(r,\mu) = \chi_{\rm line}(r)\frac{c}{\nu_0}\left[(1-\mu^2)\frac{v(r)}{r} + \mu^2\frac{{\rm d}v(r)}{{\rm d}r}\right]^{-1}$$

is the Sobolev optical depth and $\nu_0$ is the frequency at the center of each line ($\chi_{\rm line}(r)$ is the local line absorption coefficient, $\mu$ is the cosine of the angle between the ray direction and the outward normal on the spherical surface element, $v(r)$ is the local velocity, and $c$ is the speed of light). A comparison of the line acceleration of strong and weak lines evaluated with the comoving-frame method and the Sobolev technique disregarding continuum interaction is presented in Fig. 5 of Pauldrach et al. (1986), and a comparison of the comoving-frame method and the Sobolev-with-continuum technique is shown in Fig. 3 of Puls & Hummer (1988), demonstrating the excellent agreement of the two methods.

[6] For the latest update of the general method, see Pauldrach et al. (2014).

In addition, secondary effects such as the production of EUV and X-ray radiation in the cooling zones of shocks embedded in the wind and arising from the non-stationary, unstable behavior of radiation-driven winds can, together with K-shell absorption, be optionally considered (based on a parametrization of the shock jump velocity; cf. Pauldrach et al. 1994, 2001). However, they have not been included in the stellar models used to describe the ionizing sources here, since they affect only high ionization stages like He III which are not relevant for the analysis of emission line spectra of H II regions around O-type stars (see Kaschinski & Ercolano 2013).

Of course, it needs to be clarified whether the spectral energy distributions calculated by our method are realistic enough to be used in diagnostic modeling of H II regions. Although the radiation in the ionizing spectral range cannot be directly observed, the predicted SEDs can be verified indirectly by a comparison of observed emission line strengths and those calculated by nebular models (cf. Sellmaier et al. 1996; Giveon et al. 2002; Rubin et al. 2007, 2008, 2016). A more stringent test can be provided by a comparison of the synthetic and observed UV spectra of individual massive stars, which involves hundreds of spectral signatures of various ionization stages with different ionization thresholds, and covering a large frequency range: because almost all of the ionization thresholds lie in the spectral range shortward of the hydrogen Lyman edge (cf. Pauldrach et al. 2012), and the ionization balances of all elements depend sensitively on the ionizing radiation throughout the entire wind, the ionization balance can be traced reliably through the strength and structure of the wind lines formed throughout the atmosphere. In this way a successful comparison of observed and synthetic UV spectra (Pauldrach et al. 1994, 2001, 2004, 2012) ascertains the quality of the ionization balance and thus of the SEDs.

### 2.2. Numerical modelling of the H II regions

For the computation of spherically symmetric models of gaseous nebulae we have adapted the stellar atmosphere code discussed above (Pauldrach et al. 2001, 2012) to the dilute radiation fields and low gas densities in H II regions (Hoffmann et al. 2012, Weber et al. 2015). The results of this approach describe steady-state H II regions in which ionization and recombination, as well as heating and cooling are in equilibrium at each radius point.

In the stationary state, the ionization fractions of an element are given by the solution $\boldsymbol{n}^\infty$ of the equation

$$\boldsymbol{G} \cdot \boldsymbol{n}^\infty = \boldsymbol{0}, \qquad (1)$$

where $\boldsymbol{G}$ is a matrix formed from the ionization and recombination rate coefficients[7], and the components of the vector $\boldsymbol{n}^\infty$ are the number densities of successive ionization stages of the elements considered. Because the rate coefficients in general depend on the radiation field and the temperature (themselves in turn dependent on the ionization ratios), the system is solved iteratively together with the radiation field and the temperature until convergence is reached.

In our spherically symmetric models we compute the radiative transfer along a series of rays with different impact parameters to the central source, yielding the radiative intensity $I_\nu$ at every point along each ray. At any given radius shell $r$, each ray intersecting that shell corresponds to a different angle $\vartheta$ with respect to the outward normal, and the mean intensity $J_\nu(r)$ that enters into the radiative rates (photoionization and photoheating) is then obtained as $J_\nu(r) = \frac{1}{2} \int_{-1}^{1} I_\nu(r,\mu) \, d\mu$, with $\mu = \cos\vartheta$. The symmetry allows taking into account the diffuse radiation field (ionizing radiation created in the gas itself via direct recombination to the ground level) without incurring additional computational cost, and thus does not necessitate the use of simplifying approximations such as "on the spot" or "outward only".

The emission spectrum of the gas depends on both its ionization structure and its temperature, since the recombination and collisional excitation rates are temperature-dependent. Interpretation of the spectra of ionized gas thus also requires considering the microphysical processes that heat and cool the gas and thereby determine its temperature structure. In H II regions, the main heating source is the ionization of H I and He I, which increases the kinetic energy of the gas by $\Delta E = h(\nu - \nu_0)$ per ionization process ($\nu$ is the frequency of the ionizing photon and $h\nu_0$ the ionization energy). Cooling of the gas happens via recombination processes (which radiate away the kinetic energy of the electron and the ionization energy of the resulting ion, usually in several steps when recombination occurs to an excited state), radiative de-excitation of collisionally excited states, and free-free radiation (bremsstrahlung).

In H II regions, a large part of the energy of the absorbed ionizing photons is transformed into line emission in the UV, the visible, and the infrared wavelength range. These lines are either recombination lines like the Balmer lines of hydrogen, or collisionally excited lines such as the "nebulium" lines of O III at 4959 Å and 5007 Å. To compute the emission in these lines, the bound-bound transition rates among the different excitation states within the ionization stage must be calculated. Our numerical approach for computing these rates is described in detail by Hoffmann et al. (2012) and Weber et al. (2015).

### 2.3. Properties of the filtered stellar radiation

In our first set of simulations we study the emission spectrum of diffuse gas ionized by hot stars surrounded by an H II region of denser gas that attenuates and filters the stellar radiation. For this we use spherically symmetric models and assume a constant density in each of the two gas regions considered ($n_H = 10 \, \text{cm}^{-3}$ for the H II region and $n_H = 0.1 \, \text{cm}^{-3}$ for the DIG). Our model grid comprises stellar effective temperatures of 30, 35, 40, and 50 kK and uses the spectral energy distributions of the dwarf (i.e., main sequence) star models from Weber et al. (2015) with metallicities of 0.1, 0.4, and 1.0 $Z_\odot$.[8]

Our procedure involves a two-step process. In the first step, we compute models for H II regions, which we assume to be ionized by a single model star each, and iteratively adjust the radii of these (matter-bounded) H II regions until a given fraction (70%, 50%, 30%, 10%, and 5%) of the total hydrogen-ionizing photon flux of the stars is able to escape. In the second step, we use the filtered stellar radiation from these models to illuminate a homogeneous region of much thinner gas representing the DIG, which we take to be radiation-bounded in all cases. For better comparability, we scale the ionizing fluxes escaping from the H II regions such that the number of hydrogen-ionizing photons is $10^{50} \, \text{s}^{-1}$ in all cases (i.e., we compare sources with equal photon-number rates but different spectral shapes). We furthermore assume the metallicity of the DIG to be the same as that of the inner H II regions.

---

[7] The elements of $\boldsymbol{G}$ are $G_{ij} = \mathcal{P}_{ji}$ for $i \neq j$ and $G_{ii} = -\sum_{j \neq i} \mathcal{P}_{ij}$, where $\mathcal{P}_{ij}$ is the rate coefficient for the transition from ionization stage $i$ to stage $j$. The redundant system of equations is closed by replacing one of the rows by the requirement of particle conservation, $\sum_i n_i = n_{\text{tot}}$.

[8] The abundances representing solar metallicity have been taken from Asplund et al. (2009); for our models with sub-solar metallicities we have uniformly scaled the abundances of all metals with the same factor.

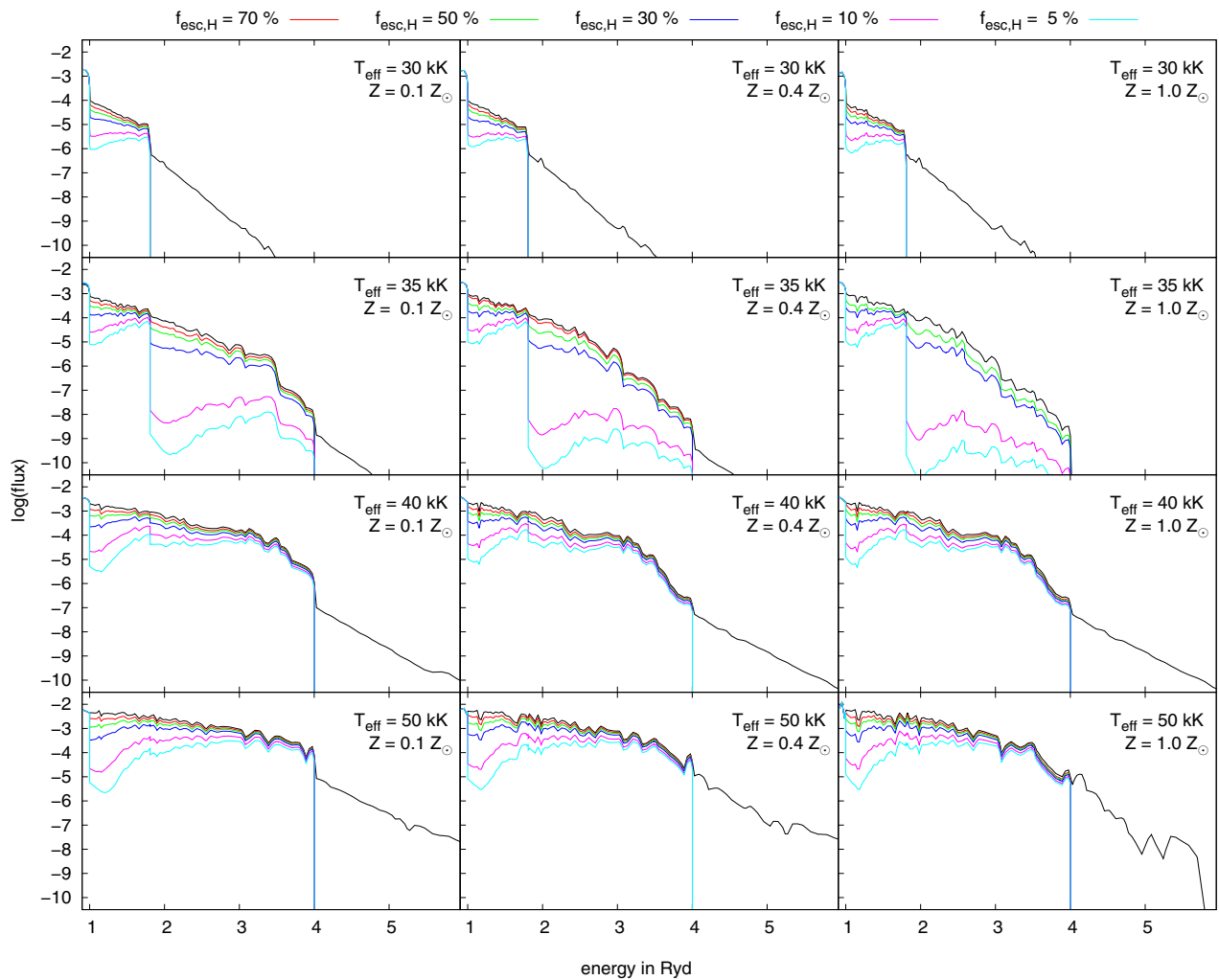

**Fig. 1.** Computed stellar SEDs (black lines) modified by homogeneous matter-bounded H II regions ($n_H = 10\,\text{cm}^{-3}$) for different escape fractions (colored lines as labelled). As sources we have used the main-sequence spectra from the hot star model grid by Weber et al. (2015). The chemical composition of the H II regions is assumed to be the same as that of the corresponding stellar atmospheres. Shown are the Eddington fluxes normalized to the stellar radius.

In our models we do not currently consider the effects of dust absorption or scattering. While dust is present in H II regions, Mathis & Wood (2005) find the absorption by silicate-based dust to be weak and – in the frequency range of ionizing radiation – almost wavelength-independent. (Thus, it may reduce the total amount of escaping ionizing radiation, but has no large effect on its spectral distribution.) They also call attention to the fact that while carbon-based dust (composed of graphite particles or polycyclic hydrocarbons) has an absorption feature that peaks at 17 eV and could therefore alter the spectral energy distribution of the ionizing radiation field, observations showing that there is almost no C-depletion from the gaseous phase indicate that the carbon-based dust in H II regions has been mostly vaporized. At present we also do not account for the different depletion of individual types of metals from the gaseous phase, as discussed, for instance, by Shields & Kennicutt (1995).

With regard to the amount of light in the optical wavelength range (including the diagnostic lines) emitted by H II regions that is scattered by dust in the DIG, the evidence is still inconclusive. While Seon & Witt (2012) find a large fraction of around 50%, Ascasibar et al. (2016) find a much smaller contribution of only about 1 to 2 percent. Thus, while we do not consider such scattering in our models and our computed line ratios therefore represent the "true" line emission by the DIG, the possibility must be kept in mind that the observed line ratios from the DIG may be contaminated to a yet undetermined degree by light not directly originating in the DIG.

In Fig. 1 we plot the SEDs $F_\star(\nu)$ of the ionizing sources, as well as the fluxes $F(\nu)$ escaping from the leaky (matter-bounded) H II regions around these sources, as a function of effective temperature, chemical composition, and the total fraction of escaping hydrogen-ionizing photons. Fig. 2 shows the spectral shape of the escape fraction $f_{\text{esc}}(\nu) = F(\nu)/F_\star(\nu)$, i.e., the "filter functions" of the H II regions.

In the energy range between the ionization edges of hydrogen and (neutral) helium, the escape fraction increases for higher photon energies because the absorption cross-section of hydrogen decreases roughly proportional to $\nu^{-3}$. However, in matter-bounded H II regions diffuse radiation from recombination processes within the gas is also emitted from the H II region into the surrounding gas. The total emission from the H II region results from the sum of the unabsorbed photons from the star and the diffuse radiation that is able to escape from the gas and is responsible for the increased emission in the energy range directly above the ionization threshold of hydrogen.

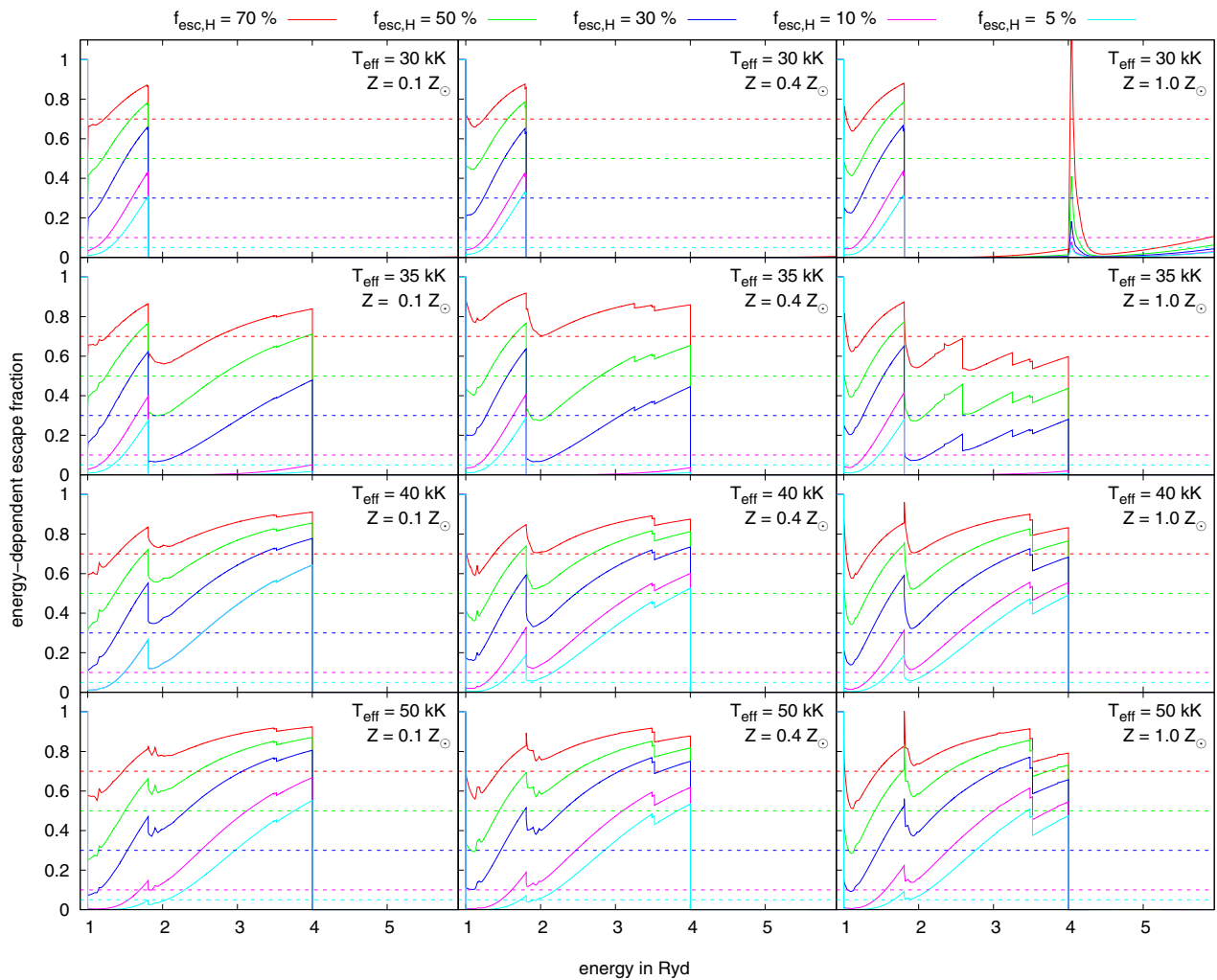

**Fig. 2.** Energy-dependent escape fractions $F(\nu)/F_\star(\nu)$ of photons above the ionization threshold of hydrogen for the fluxes shown in Fig. 1. In general, the escape fraction rises above each of the strong ionization edges due to the decreasing absorption cross sections for higher photon energies. The He II-ionizing photons are almost completely absorbed within the H II regions for all considered stellar sources; for $T_{\rm eff} = 30\,000$ K almost no He I-ionizing radiation can escape from the H II region either. For higher metallicities, in addition to the prominent ionization edges of H I, He I, and He II, there is also notable absorption by metal ions, especially of O II, Ne II, N III, and C III with ionization energies of 2.59 Ryd, 3.35 Ryd, 3.49 Ryd, and 3.52 Ryd, respectively. The dashed lines indicate the corresponding angle-averaged escape fractions for the case of extremely clumpy H II regions that would allow radiation to escape unhindered in some directions while being completely opaque to all wavelengths in others ("picket-fence" case). The spike in the escape fractions of the upper right panel is a result of the extremely low stellar He II-ionizing photon flux, which is approximately 12 orders of magnitude lower than the corresponding hydrogen-ionizing flux. Thus, even very little emission of photons by the H II region in that energy range is sufficient to exceed the flux emitted by the stellar atmosphere.

At the He I ionization edge ($h\nu_0 = 1.81$ Ryd) helium becomes an important additional absorber, so that the escape fraction again drops significantly. In the case of the 30 000 K stars, practically no ionizing radiation above the ionization edge of neutral helium can escape from the H II region. This is because the total He-ionizing photon emission rate of these stars is very small (approximately 2.5 dex smaller than the hydrogen-ionizing flux, see Weber et al. 2015), and thus the helium within the H II region is sufficient to absorb almost the entire He-ionizing radiation. This latter result is in agreement with earlier work by Hoopes & Walterbos (2003) using LTE stellar atmospheres.

For the 35 000 K stars and escape fractions of 5% and 10% of hydrogen-ionizing photons, the helium-ionizing photons are also nearly completely absorbed in the H II regions, whereas for the larger escape fractions a part of the He I-ionizing radiation can escape into the surrounding DIG. For higher stellar temperatures, the effect of the additional absorption by neutral helium becomes less important because the emission rate of helium-ionizing photons by the star grows in relation to that of hydrogen-ionizing photons. Thus, a smaller fraction of the emitted helium-ionizing photons is consumed to keep the helium in the H II regions ionized.

In the energy range above the ionization edge of He I, the escape fraction rises again as a result of ionization cross sections of both hydrogen and helium decreasing for larger frequencies. In this energy range there is also a noticeable contribution of metal ions to the absorption of the stellar radiation. The most important absorbing metal ions are O II ($h\nu_0 = 2.59$ Ryd), Ne II (3.25 Ryd), N III (3.49 Ryd), and C III (3.52 Ryd). The effect is more pronounced for larger metallicities, and for $T_{\text{eff}} \geq 40\,000$ K only the latter two ions are relevant as absorbers in our model H II regions while absorption by the former becomes negligible due to their decreasing abundance.

**Table 1.** Emission of the diagnostically important collisionally excited lines [N II] $\lambda\lambda$ 6584, 6548, [O II] $\lambda\lambda$ 3726, 3729, [O III] $\lambda\lambda$ 5007, 4959 and [S II] $\lambda\lambda$ 6716, 6731 relative to the corresponding H$\beta$ emission for matter-bounded (leaky) H II regions and the surrounding DIG. The SEDs of the ionizing sources are from Weber et al. (2015) with metallicities of 0.1 $Z_\odot$, 0.4 $Z_\odot$, and 1.0 $Z_\odot$. The H II regions have densities of $n_H = 10\,\text{cm}^{-3}$ and their outer radii adjusted to allow 70% ...5% of the ionizing photons to escape. These are used as ionizing fluxes for the DIG ($n_H = 0.1\,\text{cm}^{-3}$), scaled to $10^{50}$ ionizing photons/s for easier comparison. For the picket-fence cases the line ratios listed for the H II regions are from optically thick, radiation-bounded ($f_{esc} = 0$) models, while for the DIG the unaltered stellar SEDs (corresponding to $f_{esc} = 100\%$) are used as the ionizing fluxes (again scaled to $10^{50}$ ionizing photons/s). (Continued in Table 2.)

| $T_{\text{eff}}$ (K) | $Z$ ($Z_\odot$) | $f_{\text{esc,H}}$ | H II regions [N II] | [O II] | [O III] | [S II] | DIG [N II] | [O II] | [O III] | [S II] |
|---|---|---|---|---|---|---|---|---|---|---|
| | | | (in units of H$\beta$) | | | | | | | |
| 30 000 | 0.1 | picket fence | 0.92 | 2.1 | 0.00 | 0.38 | 0.92 | 2.0 | 0.00 | 0.33 |
| | | 70% | 0.95 | 2.3 | 0.00 | 0.21 | 0.95 | 2.1 | 0.00 | 0.38 |
| | | 50% | 0.93 | 2.2 | 0.00 | 0.27 | 1.0 | 2.6 | 0.00 | 0.25 |
| | | 30% | 0.92 | 2.1 | 0.00 | 0.31 | 1.0 | 2.4 | 0.00 | 0.35 |
| | | 10% | 0.91 | 2.1 | 0.00 | 0.35 | 1.0 | 2.5 | 0.00 | 0.27 |
| | | 5% | 0.93 | 2.1 | 0.00 | 0.37 | 0.97 | 2.4 | 0.00 | 0.22 |
| | 0.4 | picket fence | 1.7 | 2.2 | 0.00 | 0.84 | 1.7 | 2.0 | 0.00 | 0.82 |
| | | 70% | 1.5 | 1.8 | 0.01 | 0.41 | 1.8 | 2.3 | 0.00 | 0.88 |
| | | 50% | 1.5 | 1.8 | 0.01 | 0.51 | 1.9 | 2.6 | 0.00 | 0.88 |
| | | 30% | 1.5 | 1.8 | 0.00 | 0.62 | 2.0 | 3.0 | 0.00 | 0.85 |
| | | 10% | 1.6 | 2.1 | 0.00 | 0.75 | 2.2 | 3.5 | 0.00 | 0.71 |
| | | 5% | 1.6 | 2.1 | 0.00 | 0.79 | 2.2 | 3.7 | 0.00 | 0.65 |
| | 1.0 | picket fence | 1.7 | 1.1 | 0.00 | 1.1 | 1.7 | 1.1 | 0.00 | 0.72 |
| | | 70% | 1.3 | 0.79 | 0.00 | 0.64 | 1.9 | 1.3 | 0.00 | 0.97 |
| | | 50% | 1.3 | 0.77 | 0.00 | 0.73 | 2.0 | 1.5 | 0.00 | 1.0 |
| | | 30% | 1.4 | 0.83 | 0.00 | 0.83 | 2.2 | 1.8 | 0.00 | 1.0 |
| | | 10% | 1.5 | 1.0 | 0.00 | 1.0 | 2.5 | 2.3 | 0.00 | 0.92 |
| | | 5% | 1.6 | 1.1 | 0.00 | 1.1 | 2.3 | 2.2 | 0.00 | 0.77 |
| 35 000 | 0.1 | picket fence | 0.48 | 1.9 | 1.9 | 0.16 | 0.55 | 2.0 | 1.3 | 0.25 |
| | | 70% | 0.12 | 1.1 | 4.0 | 0.04 | 0.63 | 2.1 | 1.2 | 0.26 |
| | | 50% | 0.19 | 1.4 | 3.3 | 0.06 | 0.71 | 2.2 | 1.1 | 0.27 |
| | | 30% | 0.29 | 1.8 | 2.7 | 0.09 | 0.92 | 2.6 | 0.74 | 0.16 |
| | | 10% | 0.44 | 1.9 | 2.2 | 0.14 | 1.2 | 3.2 | 0.01 | 0.21 |
| | | 5% | 0.40 | 1.9 | 2.0 | 0.15 | 1.2 | 3.4 | 0.01 | 0.19 |
| | 0.4 | picket fence | 1.2 | 3.0 | 1.9 | 0.41 | 1.3 | 3.2 | 1.3 | 0.65 |
| | | 70% | 0.14 | 1.1 | 4.4 | 0.05 | 1.5 | 3.5 | 1.2 | 0.69 |
| | | 50% | 0.37 | 1.9 | 3.2 | 0.12 | 1.9 | 4.0 | 1.1 | 0.72 |
| | | 30% | 0.60 | 2.4 | 2.7 | 0.19 | 2.4 | 4.6 | 0.70 | 0.71 |
| | | 10% | 1.0 | 2.9 | 2.2 | 0.32 | 3.2 | 6.2 | 0.01 | 0.64 |
| | | 5% | 1.1 | 3.0 | 2.1 | 0.36 | 3.4 | 6.9 | 0.00 | 0.58 |
| | 1.0 | picket fence | 1.4 | 2.1 | 0.83 | 0.55 | 1.4 | 1.5 | 0.93 | 0.62 |
| | | 70% | 0.19 | 0.66 | 1.6 | 0.06 | 1.6 | 1.8 | 0.71 | 0.68 |
| | | 50% | 0.38 | 1.0 | 1.4 | 0.13 | 1.7 | 2.3 | 0.67 | 0.67 |
| | | 30% | 0.65 | 1.5 | 1.2 | 0.22 | 1.6 | 2.3 | 0.31 | 0.69 |
| | | 10% | 1.1 | 1.8 | 0.94 | 0.38 | 2.5 | 2.9 | 0.00 | 0.48 |
| | | 5% | 1.3 | 2.0 | 0.88 | 0.46 | 1.9 | 2.5 | 0.00 | 0.29 |

In none of our nebular models does a significant amount of He II-ionizing radiation ($h\nu \geq 4$ Ryd) escape, which is a consequence of the even lower stellar photon emission rates in this energy range, which are several orders of magnitude lower than the corresponding H and He I-ionizing fluxes. Even in the case of the 50 kK model star with a metallicity of 0.1 $Z_\odot$, which is the strongest emitter of He II-ionizing radiation of the stars considered here, the emission rate of He I-ionizing photons exceeds that of He II-ionizing photons by a factor of approximately 1500.

Another factor influencing the frequency-dependence of the escape fraction in real H II regions is the density structure of the filtering gas. While radiation hardening effects such as those described above are pronounced in homogeneous H II regions, the attenuation would become independent of wavelength (though strongly varying with angle) if all of the radiation were completely blocked in some directions by discrete, completely opaque clumps, and no absorbing matter at all intervened in other directions ("picket-fence" case). This latter scenario (wavelength-independent absorption by opaque clumps) and the former one (filtering of the radiation by homogeneous H II regions) represent hypothetical limiting cases. Actual H II regions (and the DIG) show a more complex density structure, with characteristics of both cases.[9] We will examine the influence of different density distributions in the context of our 3D simulations in Sect. 3 below.

**Table 2.** As Table 1, but for the 40 000 K and 50 000 K model stars as ionizing sources.

| $T_{\rm eff}$ (K) | $Z$ ($Z_\odot$) | $f_{\rm esc,H}$ | H II regions [N II] | [O II] | [O III] | [S II] | DIG [N II] | [O II] | [O III] | [S II] |
|---|---|---|---|---|---|---|---|---|---|---|
| | | | (in units of H$\beta$) | | | | | | | |
| 40 000 | 0.1 | picket fence | 0.1 | 0.50 | 5.0 | 0.09 | 0.20 | 0.85 | 3.9 | 0.15 |
| | | 70% | 0.03 | 0.24 | 6.6 | 0.02 | 0.19 | 0.80 | 3.8 | 0.17 |
| | | 50% | 0.04 | 0.32 | 6.2 | 0.02 | 0.18 | 0.72 | 4.0 | 0.18 |
| | | 30% | 0.06 | 0.41 | 5.8 | 0.04 | 0.15 | 0.59 | 4.9 | 0.18 |
| | | 10% | 0.08 | 0.48 | 5.4 | 0.05 | 0.11 | 0.38 | 5.0 | 0.19 |
| | | 5% | 0.08 | 0.49 | 5.2 | 0.07 | 0.10 | 0.29 | 5.1 | 0.20 |
| | 0.4 | picket fence | 0.32 | 1.1 | 6.7 | 0.22 | 0.71 | 2.2 | 5.2 | 0.48 |
| | | 70% | 0.06 | 0.27 | 6.5 | 0.03 | 0.71 | 2.2 | 5.7 | 0.49 |
| | | 50% | 0.09 | 0.40 | 6.3 | 0.04 | 0.67 | 2.1 | 6.6 | 0.49 |
| | | 30% | 0.14 | 0.60 | 6.5 | 0.07 | 0.61 | 2.0 | 8.2 | 0.50 |
| | | 10% | 0.27 | 0.95 | 6.8 | 0.12 | 0.46 | 1.5 | 11 | 0.53 |
| | | 5% | 0.26 | 1.1 | 6.8 | 0.15 | 0.41 | 1.3 | 13 | 0.57 |
| | 1.0 | picket fence | 0.32 | 0.64 | 3.9 | 0.27 | 0.78 | 1.4 | 3.7 | 0.61 |
| | | 70% | 0.03 | 0.07 | 2.0 | 0.02 | 0.83 | 1.5 | 4.1 | 0.63 |
| | | 50% | 0.05 | 0.09 | 2.0 | 0.03 | 0.81 | 1.6 | 5.2 | 0.66 |
| | | 30% | 0.09 | 0.16 | 2.5 | 0.05 | 0.77 | 1.5 | 7.1 | 0.70 |
| | | 10% | 0.18 | 0.37 | 3.4 | 0.11 | 0.67 | 1.5 | 12 | 0.87 |
| | | 5% | 0.23 | 0.50 | 3.7 | 0.15 | 0.62 | 1.4 | 15 | 0.97 |
| 50 000 | 0.1 | picket fence | 0.04 | 0.22 | 6.4 | 0.07 | 0.14 | 0.56 | 4.2 | 0.18 |
| | | 70% | 0.01 | 0.10 | 8.0 | 0.01 | 0.13 | 0.53 | 4.4 | 0.18 |
| | | 50% | 0.02 | 0.13 | 7.5 | 0.01 | 0.12 | 0.46 | 4.7 | 0.19 |
| | | 30% | 0.02 | 0.17 | 7.2 | 0.02 | 0.11 | 0.38 | 5.0 | 0.19 |
| | | 10% | 0.03 | 0.29 | 6.7 | 0.03 | 0.09 | 0.27 | 5.2 | 0.23 |
| | | 5% | 0.03 | 0.32 | 7.5 | 0.04 | 0.06 | 0.17 | 5.6 | 0.20 |
| | 0.4 | picket fence | 0.11 | 0.45 | 10 | 0.16 | 0.43 | 1.4 | 9.0 | 0.34 |
| | | 70% | 0.02 | 0.18 | 9.2 | 0.01 | 0.42 | 1.4 | 9.9 | 0.54 |
| | | 50% | 0.03 | 0.15 | 9.4 | 0.02 | 0.39 | 1.4 | 11 | 0.56 |
| | | 30% | 0.04 | 0.23 | 10 | 0.04 | 0.37 | 1.2 | 13 | 0.60 |
| | | 10% | 0.07 | 0.36 | 11 | 0.07 | 0.33 | 0.96 | 17 | 0.73 |
| | | 5% | 0.08 | 0.41 | 11 | 0.09 | 0.32 | 0.85 | 18 | 0.84 |
| | 1.0 | picket fence | 0.16 | 0.48 | 7.2 | 0.24 | 0.60 | 1.5 | 6.9 | 0.71 |
| | | 70% | 0.01 | 0.05 | 3.3 | 0.01 | 0.61 | 1.6 | 8.1 | 0.75 |
| | | 50% | 0.02 | 0.07 | 3.5 | 0.02 | 0.60 | 1.6 | 10 | 0.80 |
| | | 30% | 0.04 | 0.11 | 4.4 | 0.03 | 0.60 | 1.6 | 13 | 0.90 |
| | | 10% | 0.07 | 0.24 | 6.1 | 0.07 | 0.58 | 1.6 | 21 | 1.2 |
| | | 5% | 0.09 | 0.32 | 6.8 | 0.10 | 0.58 | 1.6 | 21 | 1.2 |

We approximate the "picket-fence" case by increasing the radii of the H II regions until they become radiation-bounded, while also using the unfiltered SEDs of the model stars as the ionizing fluxes for the DIG (again scaled to $10^{50}$ ionizing photons per second).

### 2.4. Line ratios in the matter-bounded (leaky) H II regions

In Tables 1 and 2 we show the predicted luminosities of collisionally excited line radiation (relative to H$\beta$) both of the matter-bounded H II regions and of the DIG irradiated by the ionizing fluxes escaping from such H II regions.

For the matter-bounded H II regions we find that the ratio of the lines emitted by the singly ionized atoms of N, O, and S relative to the H$\beta$ emission increases for lower escape fractions. The reason for this behavior is that these ions occur predominantly in those parts of the H II regions where the optical depth to the emitting source is large, i.e., near the Strömgren radius of a radiation-bounded region. The increase is small for the H II regions ionized by the 30 000 K model stars, because due to their relatively low temperature, these stars emit only a very small number of photons capable of creating doubly ionized states of those elements (cf. Table 4 of Weber et al. 2015). Thus for the 30 000 K stars the singly ionized atoms are abundant even in the inner parts of the H II regions.

---
[9] For instance, Wood et al. (2005) have modeled fractal density structures, where overdense regions exist at different size scales, which are, however, not necessarily optically thick.

Another factor that contributes to enhanced emission from the singly ionized species for lower escape fractions is that the strongest radiation hardening arises for the largest optical depths for ionizing radiation, which occur only for low escape fractions. The hardening results in increased heating rates and gas temperatures, and, consequently, an increased emission of collisionally excited lines in the optical wavelength range. In the case of the metal-poor H II regions with $Z = 0.1 Z_\odot$ the latter effect is compensated by the enhanced cooling via collisional excitation of neutral hydrogen, which is more abundant in the outer parts of H II regions with low escape fraction.[10]

The [O III]/H$\beta$ ratio as a function of the escape fraction strongly depends on the effective temperature of the ionizing sources. For the 30 000 K stars, this ratio is $\leq 10^{-3}$ for all metallicities and all escape fractions, due to the very low O II-ionizing fluxes of stars at that temperature, which lie more than 4 orders of magnitude below the hydrogen-ionizing fluxes (Weber et al. 2015). For an effective temperature of 35 000 K the [O III]/H$\beta$ ratio drops for lower escape fractions, as the presence of O III ions is limited to a smaller volume around the source compared to ionized hydrogen responsible for the H$\beta$ recombination radiation. For the model stars with 40 000 K and 50 000 K the [O III]/H$\beta$ ratio as a function of the escape fraction is metallicity dependent. The ratio decreases for lower escape fractions in the case of a (nebular) metallicity of $0.1 Z_\odot$ whereas it increases for the metallicities of $0.4 Z_\odot$ and $1.0 Z_\odot$. This is caused by the different radius-dependence of the temperature in the H II region. The temperature – and consequently the excitation of optical O III lines – increases with radius due to the radiation hardening for the metallicities of $0.4 Z_\odot$ and $1.0 Z_\odot$. In contrast, the temperature (slightly) decreases at larger radii for $Z = 0.1 Z_\odot$, which is caused by the cooling by collisionally excited neutral hydrogen, which is more abundant at larger optical depths in the outer parts of the ionized region. Consequently, the emission of the collisionally excited lines (relative to H$\beta$) of those ions still present near the Strömgren radius is enhanced in comparison to that of matter-bounded H II regions.

## 2.5. Line ratios in the DIG

For the DIG regions we find that the strength of the [N II] and [O II] emission lines increases relative to H$\beta$ for decreasing escape fractions where the effective temperature of the source is 30 000 K or 35 000 K and the metallicity is $0.4 Z_\odot$ or $1.0 Z_\odot$. This is caused by the larger temperatures due to the radiation hardening effect. In contrast, the [S II]/H$\beta$ emission line ratio overall tends to decrease for DIG ionized by radiation escaping from the optically thicker H II regions, but the line ratio as a function of the escape fraction is often non-monotonic and reaches it highest values not in the picket-fence case, but for DIG irradiated by radiation passing through H II regions with escape fractions of 30% to 50%. This is due to two opposing effects on the abundance of singly ionized sulfur in the ionized volume. Since the ionization energy of singly ionized sulfur (23.3 eV) is slightly lower than that of neutral helium (24.6 eV), the fraction of radiation in the interval between these ionization energies relative to the total hydrogen-ionizing radiation rises as a result of radiation hardening and can lead to a reduction of the number of S II ions in the DIG ionized by the filtered radiation. On the other hand, lower escape fractions in H II regions around the model stars with $T_{\text{eff}} \leq 35\,000$ K also lead to a more pronounced He I/He II absorption edge, and consequently fewer photons with energies above the corresponding ionization threshold are available to ionize the remaining S II. The latter effect is mainly relevant for the 35 000 K model stars, as the emission of He I-ionizing radiation from the 30 000 K model stars is very low, so that the contribution of radiation in that wavelength range to the total S II-ionizing radiation is negligible.

---

[10] Due to the less efficient cooling by metal ions for low values of $Z$, the temperature of the ionized gas is higher for lower metallicities. The higher temperature in turn increases the probability of collisional excitation of neutral hydrogen, where the energy gap between the ground state and the lowest excited level is considerably larger than for the metal ions (cf. the discussion by Weber et al. 2015).

For the radiation sources with effective temperatures of 30 000 K, almost no O II-ionizing radiation can escape from the H II regions, so that there is no [O III] line radiation from the DIG. The same is true for the H II regions around the 35 000 K model stars with small escape fractions of 10% or 5%. For larger escape fractions the [O III]/H$\beta$ ratio in the DIG rises quickly, with the largest ratio reached for the largest escape fractions.

In contrast, for the higher stellar temperatures of 40 000 K and 50 000 K the [O III]/H$\beta$ ratio increases for decreasing escape fractions, as for these sources O II-ionizing photons can leave the H II regions even for small escape fractions of 10% or 5%. Actually, the fraction of O II-ionizing photons of all hydrogen-ionizing photons that can escape from the H II region is larger than the corresponding fraction of ionizing photons originally emitted by the sources, and the ratio increases for smaller escape fractions due to radiation hardening. For instance, we find an increase of approximately 0.6 dex for model stars with an effective temperature of 40 000 K and an escape fraction of 10%. Additionally, the radiation hardening leads to increased temperatures in the DIG. The radiation hardening also leads to decreasing [N II] and [O II] emission in the DIG around H II regions with small escape fractions for stellar temperatures of $T_{\text{eff}} \geq 40\,000$ K. In this case the effect of the larger fraction of photons able to strip a further electron from these ions is more important than the temperature increase due to radiation hardening.

The situation is different for S II, where the emission increases for lower escape fractions. The ionization energy of neutral sulfur lies below the ionization threshold of hydrogen, while the energy required for doubly ionized sulfur lies below the ionization threshold of neutral helium. Around hot O stars with strong He I-ionizing fluxes, S II ions therefore exist predominantly in the neutral phase of the medium. At the same time, electrons are required for collisional excitation processes. Most of the S II emission therefore takes place at the interface region between the ionized and the surrounding neutral phase, which is broadened by the radiation hardening as higher photon energies lead to reduced ionization cross-sections, allowing the photons to penetrate deeper into the neutral medium near the Strömgren radius.

In addition to the DIG models where the spectral energy distribution of the ionizing radiation is modified by dense H II regions in the stellar environment, we have also considered the case where the shape of the ionizing spectrum remains unchanged, and which corresponds to either of the following two scenarios: (a) as a limit case for very high escape fractions, and (b) in situations where the only absorbers are dense clumps that completely absorb the entire ionizing radiation and coexist with "channels" devoid of gas through which photons can escape unhindered from the star (see above). The results of these simulations are presented in Tables 1 and 2 as the "picket fence" case. In general, the simulated line ratios in the "picket-fence" case continue the trend of the line ratios for higher escape fractions into the DIG at the respective temperature and metallicity. Even if the stellar radiation from the $T_{\rm eff}$ = 30 000 K model stars is not filtered, the production of photons above the ionization edge of O II is too low to cause a significant amount of [O III] emission from the DIG. For the models with $T_{\rm eff}$ = 35 000 K sources, the emission lines from N II and O II are weaker than for the filtered spectra, which is caused by the lack of radiation hardening and the resulting cooler temperatures of the gas. The O III line, in contrast, is stronger than with filtered radiation, where for $T_{\rm eff}$ = 35 000 K a large fraction of the photons above the He I-ionization edge are consumed within the H II regions.

In the stellar temperature range of $T_{\rm eff} \geq 40\,000$ K the trend we have observed for increasing escape fractions in the "filtering" case (i.e., weaker absorption), namely a reduced emission of the O III line, continues. There is, however, no further significant increase in the emission from singly ionized nitrogen and oxygen.

## 2.6. Comparison with observations and other theoretical work

As noted in the introduction, observations of the diffuse ionized gas in the Milky Way indicate a softer ionizing spectrum in the DIG compared to typical H II regions. For example, measurements by Madsen et al. (2006) of the H$\alpha$ and He I $\lambda$ 5876 recombination lines as well as the collisionally excited lines [N II] $\lambda$ 6583, [N II] $\lambda$ 5755, [O III] $\lambda$ 5007, and [S II] $\lambda$ 6716 emitted by the warm interstellar medium of the Milky Way (in the direction of the Orion-Eridanus Bubble) show line ratios (in terms of energy) of [O III]/H$\alpha$ < 0.15.[11] In combination with the weak He I $\lambda$ 5876 recombination line emission (He I/H$\alpha$ ≤ 0.015) and the strong emission of [N II] and [S II] ([N II]/H$\alpha$ ~ 0.16...0.43; [S II]/H$\alpha$ ~ 0.13...0.29) this shows that there is only a small fraction of ions that need photons above the ionization edge of He I to be formed, indicating that the DIG is ionized by a soft radiation field. A comparison with Tables 1 and 2 shows that such conditions can be attained by ionizing radiation that is predominantly emitted by relatively cool O stars ($T_{\rm eff} \lesssim 35\,000$ K).

For example, for our models using an unaltered spectrum of a 35 000 K star we find an [N II]/H$\beta$ ratio of 1.3 for a metallicity of 0.4 $Z_\odot$ and 1.4 for a metallicity of 1.0 $Z_\odot$. By contrast, the radiation hardening and the removal of photons able to ionize He I lead to considerably increased [N II]/H$\beta$ ratios of up to ~3.4. While these clearly exceed the values in the numerical model DIG by Barnes et al. (2014) (where the ionization field is also described by the radiation field of a 35 000 K star), such large values have indeed been observed in the diffuse ionized gas associated with the Perseus arm by the WHAM Survey (Haffner et al. 1999, Haffner et al. 2003, and Madsen et al. 2006; collected data have been presented by Barnes et al. 2014). Regarding the [S II] emission, our simulation results of [S II]/H$\beta$ ~ 0.6 are considerably lower than the simulation results from Barnes et al. 2014

(who found values on the order of 2.0), but are in better agreement with the observational values both in the DIG associated with the Perseus arm and the interarm DIG. The results for the DIG models using a 30 000 K SED are similar, but the [N II]/H$\beta$ ratio is less dependent on the escape fractions of the filtering H II regions and remains below the H$\alpha$/H$\beta$ ratio. This is not surprising, because in both cases most of the ionized hydrogen volume predominantly contains ionization stages requiring photons below the He I ionization edge to be created. Thus, the results are in agreement with the low [O III]/H$\beta$ ratios found in the DIG of the Milky Way.

Our results are also in agreement with previous work that connects the ionization structure of the DIG with the radiation from stars at the cooler end of the O star temperature range, like that of Reynolds & Tufte (1995) who called attention to earlier observations by Torres-Peimbert et al. (1974), who found that the fraction of hot massive stars that are no longer embedded in H II regions rises for late-type O stars and B stars, i.e, stars with relatively low temperatures and, consequently, low fluxes above the ionization edge of neutral helium. This would signify higher escape fractions for the radiation from these objects and thus the ionization structure of the DIG would be dominated by the ionizing radiation of these objects. However, Reynolds & Tufte (1995) also point out that such objects contribute only about one fifth of the total stellar hydrogen-ionizing radiation, i.e., most of the radiation is contributed by hotter stars. To ionize the DIG, most of the ionizing radiation from these cooler stars would therefore have to remain unabsorbed by gas in their immediate environment.

For comparison with the DIG, Madsen et al. (2006) have also observed "classical" H II regions around hot stars, where they have found strongly varying [O III]/H$\alpha$ ratios ranging from 0.025 to 0.45, i.e., the highest ratios found in the sample of H II regions exceed the highest values found in the sample of DIG subregions by a factor of approximately four.[12]

A relatively weak emission of doubly ionized oxygen ([O III]/H$\alpha$ ~ 0.20) has also been found in the diffuse ionized gas of M 31 by Greenawalt et al. (1997), although there the discrepancy in the [O III] emission (normalized to the recombination lines) between DIG and H II regions is smaller than in the Milky Way. These line ratios shows the best agreement with our model DIG around the 35 000 K stars, but indicate a slightly harder ionizing spectrum than found in the nearby DIG of the Milky Way. We note that, as can be seen from Tables 1 and 2, even a slight variation in the effective temperatures of the ionizing sources or the escape fraction may drastically change the [O III] emission.

In contrast, an ionizing spectrum dominated by stars with effective temperatures of $T_{\rm eff}$ = 40 000 K or above would cause O III to become the strongest emission line in the optical wavelength range. The radiation hardening and the resulting higher temperatures in the DIG would lead to enhanced [O III] ratios compared to the dense H II regions directly surrounding the ionizing stars, which contradicts the observational results obtained by Greenawalt et al. (1997) and Madsen et al. (2006).

However, the dominance of soft ionizing sources is not consistent with observations that show that at least in some galaxies the [O III]/H$\beta$ ratio rises in the dilute gas at larger distances from the hot stars near the galactic disk (Otte et al. 2001, Collins & Rand 2001), because for scenarios with radiation-bounded H II regions around O stars with effective temperatures of $T_{\rm eff} \lesssim$

---

[11] The H$\alpha$/H$\beta$ ratio is approximately 2.9 and changes only weakly as a function of temperature and electron density (cf. Hummer & Storey 1987, Osterbrock & Ferland 2006). Thus, in these regions, the strength of the [O III] emission is less than half of the H$\beta$ emission we use as a reference in Tables 1 and 2.

[12] For an H II region of gas ionized by a much hotter DO white dwarf (whose temperature was determined to be around $T_{\rm eff} \approx 120$ kK by Mahsereci et al. 2012) they found a much larger ratio of [O III]/H$\alpha$ ~ 2.6.

35 000 K the harder radiation above the helium ionization threshold is restricted to the inner parts of the ionized volume, as can be seen from the results in Table 1. The situation may be different if not only the H II regions surrounding the stars, but also the DIG is matter-bounded, so that even for a softer spectral energy distribution of the ionizing sources radiation hardening may outweigh the absorption of photons above the He I ionization threshold.

Alternatively, these observational results could be explained by additional sources with high effective temperatures that are not as concentrated towards the galactic disk as massive O stars, for example the evolved stars (central stars of planetary nebulae and white dwarfs) proposed by Sokolowski & Bland-Hawthorn (1991) and Flores-Fajardo et al. (2011). Integral field observations by Lacerda et al. (2018), who find the DIG in galaxies with low current star formation rates to have higher [O III]/H$\beta$ ratios, support the idea that evolved stars may play an important role in the ionization of the more dilute parts of the DIG. In view of these results it appears unlikely that the rising [O III]/H$\beta$ ratios are mainly the result of hardening of the radiation of hot massive stars. Our results in the previous sections, which are quantified in Table 2, show that for stellar effective temperatures of $T_{\text{eff}} \gtrsim 40\,000$ K, He I-ionizing photons are able to leave the inner H II regions even for low escape fractions of the total ionizing radiation, so that the effect of radiation hardening in the energy range between the ionization thresholds of He I and He II leads to an enhanced [O III]/[O II] ratio in the DIG. But hot massive stars are extremely rare in early-type galaxies (ellipticals and gas-depleted S0 disk galaxies with little ongoing star formation), while there still exists a population of evolved low-mass hot stars. So an ionization by these objects appears to be more plausible.

## 3. Simulations of the diffuse ionized gas in 3D

In star-forming galaxies there are typically multiple regions containing hot stars that may contribute to the ionization of the DIG. Simulations of the DIG should therefore be able to simultaneously account for several sources located at different positions and embedded in an inhomogeneous interstellar gas. To realize this aim we have applied our 3D radiative transfer code to model the DIG above a spiral arm. The underlying methods of this code are described in detail by Weber et al. (2013) (geometrical aspects of the ray tracing, computation of the ionization structure of hydrogen and helium) and Weber et al. (2015) (ionization structure of metals, computation of the temperature structure of the gas, and computation of the line emission). As in the spherically symmetric models, we assume the ionized gas to exist in two forms, H II regions in the immediate environment of the stars (unresolved in these simulations) and the dilute and extended DIG.

It is not our aim here to exactly reproduce the line ratios of any particular spiral galaxy. Instead, we want to know how varying the inhomogeneity in the DIG affects the ionization structure and the line emission of the diffuse gas. Our approach is complementary to that of Wood et al. (2004, 2005), who concentrated on simulating the escape of ionizing radiation from H II regions in the stellar environment, whereas we focus on inhomogeneities within the diffuse ionized gas itself. Here our approach is similar to that of Wood et al. (2010), Barnes et al. (2014), and Barnes et al. (2015), who performed photoionization simulations of the DIG based on a density structure obtained from hydrodynamical simulations, or, in the latter case, a fractal density structure as well, while we choose a simpler description, but study how the clumpiness of the DIG affects its observable properties.

### 3.1. Method: 3-dimensional radiative transfer based on ray tracing

The difficulty in modelling the DIG is related to the inherent lack of symmetry of the problem; the presence of multiple ionization sources coupled with the inhomogeneous gas distribution can create a complex 3-dimensional pattern in the radiation field. To compute that pattern numerically a suitable discretization scheme is required. The simplest but still fully general such method is to divide up the simulation volume into a Cartesian grid of cells, where the gas density may vary from cell to cell but the ionization structure and density within each cell is taken to be homogeneous. The accuracy and detail level of the simulation may be then chosen by selecting an appropriate cell size.

Because the spatial extent of the ionized region exceeds stellar radii by typically more than six orders of magnitude, the sources of ionizing radiation can be considered to be point-like, and the radiative transfer can be computed along rays emanating from these points. Limiting the placement of the sources to the centers of cells results in no significant further loss of detail beyond the already effected discretization of the volume into cells, but allows conserving computational resources in calculating the radiation field, because then the intersections of the rays with the cell boundaries relative to the source are the same for all sources and need to be computed only once; the results are stored and then accessed via a lookup table. The rays themselves are distributed evenly in all directions around a source (using the tessellation scheme of Górski et al. 2005), each ray representing a fixed solid angle $\Delta\Omega = 4\pi/N_{\text{rays}}$ and thereby a given fraction $L_{\text{ray}} = L/N_{\text{rays}}$ of the source's luminosity $L$. The geometric dilution of the radiation field is automatically provided for by the decreasing number of rays intersecting each cell with increasing distance from the source.

Each source may have a different luminosity and spectral energy distribution, and the radiative transfer is computed for a number of frequency points (typically 100 . . . 1000) sufficient to describe the energy distribution of the ionizing flux. We account for the diffuse radiation field (the ionizing radiation emitted by recombination processes within the gas) by applying the case-B approximation for hydrogen and helium, i.e., the assumption that the photons emitted by recombination directly into the ground state are re-absorbed locally. (In the context of our 3D radiative transfer the accuracy of this approximation is quantitatively discussed by Weber et al. 2013.)

As a further step in our simulation procedure the time-dependent ionization fractions $n \equiv n(\mathbf{r}, t)$ of all ionization stages $i$ of the elements considered are calculated on the basis of the rate equation system

$$\frac{\mathrm{d}}{\mathrm{d}t} n_i(t) = \sum_{j \neq i} \mathcal{P}_{ji} n_j(t) - \sum_{j \neq i} \mathcal{P}_{ij} n_i(t), \qquad (2)$$

which describes the rate of change of the number density of all ionization stages $i$. The rate coefficients $\mathcal{P}_{ij}$ are the sums of the radiative ($\mathcal{R}_{ij}$) and the collisional ($C_{ij}$) transition rates. Because the strengths of emission lines depends on the temperature, the heating and cooling processes that determine the temperature structure of the gas must also be accounted for.

In the present work we restrict ourselves to the computation of the special case of the stationary state[13] for the temperature and ionization structure (i.e., the case where the left-hand side of Eq. 2 becomes zero), which is consistent with our assumption (see below) of continuous ongoing star formation. For otherwise fixed conditions (density structure of the gas, spectral energy distributions of the ionizing sources), the stationary equilibrium case would be reached for $t \to \infty$.[14]

To determine this equilibrium, it is necessary to compute the ionization and recombination rates and the thermal equilibrium. The photoionization rate coefficients $\mathcal{R}_{ij}(m)$ from a lower stage $i$ to an upper stage $j$ are calculated for a grid cell $m$ as

$$\mathcal{R}_{ij}(m) = \int_{\nu_{ij}}^{\infty} \frac{4\pi a_{ij}(\nu)}{h\nu} J_\nu(m) \, d\nu, \qquad (3)$$

where $J_\nu(m)$ is the mean intensity in the cell, $\nu_{ij}$ the threshold frequency for the corresponding ionization edge, and $a_{ij}(\nu)$ the frequency-dependent ionization cross-section. Similarly, the radiative heating rates, i.e., the gain of thermal energy per unit volume due to ionization of state $i$ are computed as

$$\Gamma_{ij}(m) = n_i(m) \int_{\nu_{ij}}^{\infty} \frac{4\pi a_{ij}(\nu)}{h\nu} J_\nu(m) h(\nu - \nu_{ij}) \, d\nu, \qquad (4)$$

where $n_i(m)$ is the number density of ionization stage $i$ in cell $m$.

The link between the radiative transfer and the photoionization and heating rates is the mean intensity $J_\nu(m)$, which can be computed by making use of the fact that the number of ionization processes within a grid cell must equal the number of ionizing photons absorbed in the grid cell (see Weber et al. 2013). This leads to the expression

$$J_\nu(m) = \frac{1}{4\pi V(m)} \sum_{\substack{\text{ray} \\ \text{segments}}} \frac{L^{\text{inc}}_{\text{ray},\nu}(1 - e^{-\tau_\nu})}{\chi_\nu(m)}, \qquad (5)$$

where $V(m)$ is the volume of the cell, $L^{\text{inc}}_{\text{ray},\nu}(m)$ the frequency-dependent incident luminosity of a ray entering the cell (obtained from the radiative transfer solution), $\tau_\nu(m)$ the optical depth of the ray segment in the cell, and $\chi_\nu(m)$ the opacity of the material. To obtain the mean intensity, the contributions of all rays crossing the grid cell are added up.

---

[13] We have already used our method to study time-dependent effects regarding the reionization of the universe and expanding ionization fronts in H II regions (Weber et al. 2013, 2015).

[14] However, it should be noted that the recombination and cooling time scales of the DIG are on the order of a few Myr, i.e., on the same time scale as the lifetime of the most massive stars. (The recombination timescale is $t_{\text{rec}} = 1/(\alpha_B n_H)$, which is $\sim 1.2$ Myr for a hydrogen number density of $n_H = 0.1 \, \text{cm}^{-3}$ and a temperature of $T = 10\,000$ K with a corresponding case-B recombination coefficient of $\alpha_B(T) = 2.6 \times 10^{13} \, \text{cm}^3 \, \text{s}^{-1}$ (Osterbrock & Ferland 2006).) Thus the ability of our numerical method to account for time-dependent effects will be important in future work that considers transient objects like starbursts or supermassive stars (cf. Pauldrach et al. 2012, Weber et al. 2013) with lifetimes similar to the recombination timescale. In this case the ionization structure of the diffuse ionized gas is – even if photoionization is assumed as the only significant energy source of the DIG – not only affected by the properties of the current population of ionizing sources, but also by their recent history.

The recombination rate coefficients in grid cell $m$ are $\mathcal{R}_{ji}(m) = \alpha_j(T(m)) \, n_e(m)$, where $n_e$ and $T$ are the electron density and temperature in the cell and $\alpha_j(T)$ is interpolated from tabulated, temperature-dependent recombination coefficients containing both radiative and dielectronic contributions. For hydrogen and helium we use the tables of Hummer (1994) and Hummer & Storey (1998); for the metal ions we use the fit functions given by Aldrovandi & Pequignot (1973), Aldrovandi & Pequignot (1976), Shull & van Steenberg (1982), Nussbaumer & Storey (1983), and Arnaud & Rothenflug (1985), as compiled by D. A. Verner.[15]

To compute the temperature, we consider ionization heating and recombination cooling by hydrogen and helium, as well as free-free cooling and cooling via collisional excitation and radiative de-excitation of bound states. The latter includes cooling through the collisionally excited optical lines of N II, N III, O II, O III, O IV, Ne III, S II, and S III and the fine-structure lines of N II, N III, O III, O IV, Ne II, Ne III, S III, and S IV.

The computational effort required to create 3D models of the photoionized gas considerably exceeds the effort for spherically symmetric models. For instance, in our three-dimensional simulations the number of volume elements is approximately $10^6$ and we consider the interaction of the matter of the DIG with 50 sources of ionization, whereas in the spherically symmetric models there is a single source of ionization and the temperature and ionization structure is computed for just 101 radius points. To keep the run time and memory requirement of the 3D simulations at a feasible level, we have utilized a simplified modelling of the occupation numbers in the ionized gas, where we use the "nebular approximation", i.e., we assume that the time scales for all processes that require the interaction between two particles, like ionization (requiring interaction between an atom/ion and a photon) and recombination and collisional excitation (both requiring interaction between an atom/ion and an electron) are long compared to the times scale for the relaxation of an excited atom/ion into the ground state. Thus, practically all atoms/ions that undergo interactions with photons or electrons are in their ground state and only transitions from the ground state need to be considered. As a consequence, only the abundances of the different ionization stages have to be computed explicitly, whereas in the spherically symmetric case we explicitly compute the occupation numbers for up to 50 excited levels per ion.

For the recombination processes, which can involve various excited intermediate stages during the recombination from the ground state of the higher to the ground state of the lower ionization stage, the 3D method is based fit formulae that describe the total recombination rate, i.e., the sum of the recombination rates to all levels.

For the densities considered in this work ($n_e \leq 10 \, \text{cm}^{-3}$) the nebular approximation is a good representation of the actual situation in the gas, as even for forbidden transitions like the $^1D \to {}^3P$ transitions of N II and O III, the fine-structure transitions between the $^3P$ sublevels of these ions, and the $^2D \to {}^4S$ transitions of O II and S II, the spontaneous radiative relaxation is much more probable than a two-particle interaction in the excited upper state.[16]

---

[15] http://www.pa.uky.edu/~verner/rec.html

[16] For example, the critical electron densities, where collisional de-excitation of an excited state is as probable as spontaneous radiative decay, are $6.8 \times 10^5 \, \text{cm}^{-3}$ for the $^1D_2$ and $5.1 \times 10^2 \, \text{cm}^{-2}$ for the $^3P_1$ level of O III (Osterbrock & Ferland 2006).

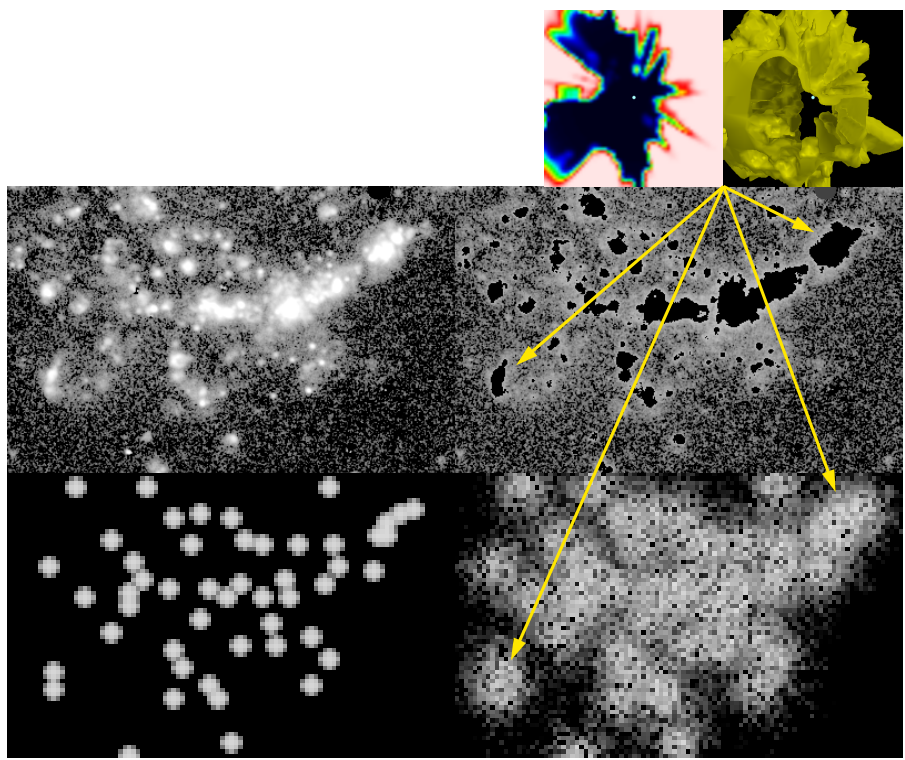

**Fig. 3.** Comparison between H$\alpha$ observations of H II regions and the DIG in the face-on galaxy M 81 by Greenawalt et al. (1998) (middle row) and our simulation results (bottom row) of the ionization of the DIG by radiation escaping from the H II regions in the environment of the radiation sources. We find a reasonable qualitative agreement regarding both the dense and relatively intense H II regions around the star-forming regions on the left-hand side of the image and the diffuse ionized gas on the right-hand side of the image. (The simulation results presented in this image have been computed for a combination of microclumping and macroclumping with a clumping factor of $f_{cl} = 5$ each.) The brightness values in these images as well as in Fig. 4 represent an emission measure range between 1 and 750 pc cm$^{-6}$ (logarithmic scaling). Observational images: ©AAS. Reproduced with permission.

### 3.2. Simulations of the DIG around a spiral arm

With our 3D simulations, we study the ionization ratios in the DIG above and below a section of the galactic disk corresponding to a spiral arm. In contrast to our parameter study in the previous section using spherically symmetric models and single-star spectra, where we examine how the line ratios vary with metallicity and hardness of the ionizing spectrum, the intent of the simulations here is to examine how the ionization structure of the DIG varies as a function of small-scale inhomogeneities in the gas. For this purpose we choose a more realistic SED from a stellar population synthesis model[17] as a common ionizing spectrum for the different clumping scenarios we consider.

In each of our simulations, the stellar population that emits the ionizing radiation is represented by 50 sources, whose positions roughly resemble the distribution of luminous H$\alpha$ emitters in a spiral arm of M 81 observed by Greenawalt et al. (1998). The distribution of distances of the sources from the galactic plane follows an exponential law with a scale height of 90 pc (Bahcall & Soneira 1984). To determine the ionizing photon emission rate of the sources, we take the ionizing luminosity predicted by the stellar population synthesis model for the star formation rate of a typical star-forming galaxy and scale it to the fraction of the galactic disk represented by our simulation volume, then divide it by the number of sources in our simulation. The resulting hydrogen-ionizing flux emitted by each source is $7.1 \times 10^{50}$ photons per second, and the emission rate of photons able to ionize He I and He II is $1.1 \times 10^{50}$ s$^{-1}$ and $2.1 \times 10^{46}$ s$^{-1}$, respectively.

Specifically, these values were obtained as follows. We assume a star-forming disk galaxy with a continuous star formation rate of approximately $1\,M_\odot$ yr$^{-1}$ like the Milky Way (Robitaille & Whitney 2010) or M 81 (Gordon et al. 2004) and a scale length of the disk of approximately 2 kpc (Porcel et al. 1998), which for a Kroupa IMF with an upper mass limit of 100 $M_\odot$ in total emits $1.5 \times 10^{53}$ hydrogen-ionizing photons per second. The ionizing flux was then adapted to a simulated volume of $(5\,\text{kpc})^3$, assuming that the center of the simulated volume has a distance of 7.5 kpc from the galactic center (similar to the distance between the galactic center and the position of the sun) and the result multiplied by 10, as we assume an enhanced star formation rate in a spiral arm of a star-forming galaxy. This star formation rate was then used as input to the population synthesis code Starburst99 (Leitherer et al. 1999, Leitherer et al. 2010, Leitherer et al. 2014, Vazquez et al. 2005), where we have chosen the recent Geneva 2012 stellar evolutionary tracks (Ekström et al. 2012, models without rotation) at solar metallicity. The luminosity computed for the simulation volume is then distributed evenly[18] among the 50 sources we have assumed in our simulations.

In a first benchmark simulation we assume that the entire (unfiltered) radiation is absorbed by H II regions with a hydro-gen number density of $2\,{\rm cm}^{-3}$. Even with this (for H II regions) comparatively low density, the radii of the ionized regions do not exceed 150 pc. These H II regions are shown in the bottom left panel of Fig. 3, where we compare the visual appearance of both the H II regions (left-hand side) and of the DIG (right-hand side) from our 3D model with an observation by Greenawalt et al. (1998). Among our simulations of the DIG we find the best agreement to the observations in a model that combines clumping at various scales (see below).

---

[17] In this regard our approach differs from that of Barnes et al. (2014, 2015), who have chosen the spectral energy distribution of a 35 000 K model star to represent the ionizing radiation field. In Sect. 3.2.1 we show a comparison model with such a softer ionizing spectrum.

[18] We currently do not consider a luminosity function of H II regions as described, for instance, by Kennicutt et al. (1989), Oey & Clarke (1998), and Liu et al. (2013).

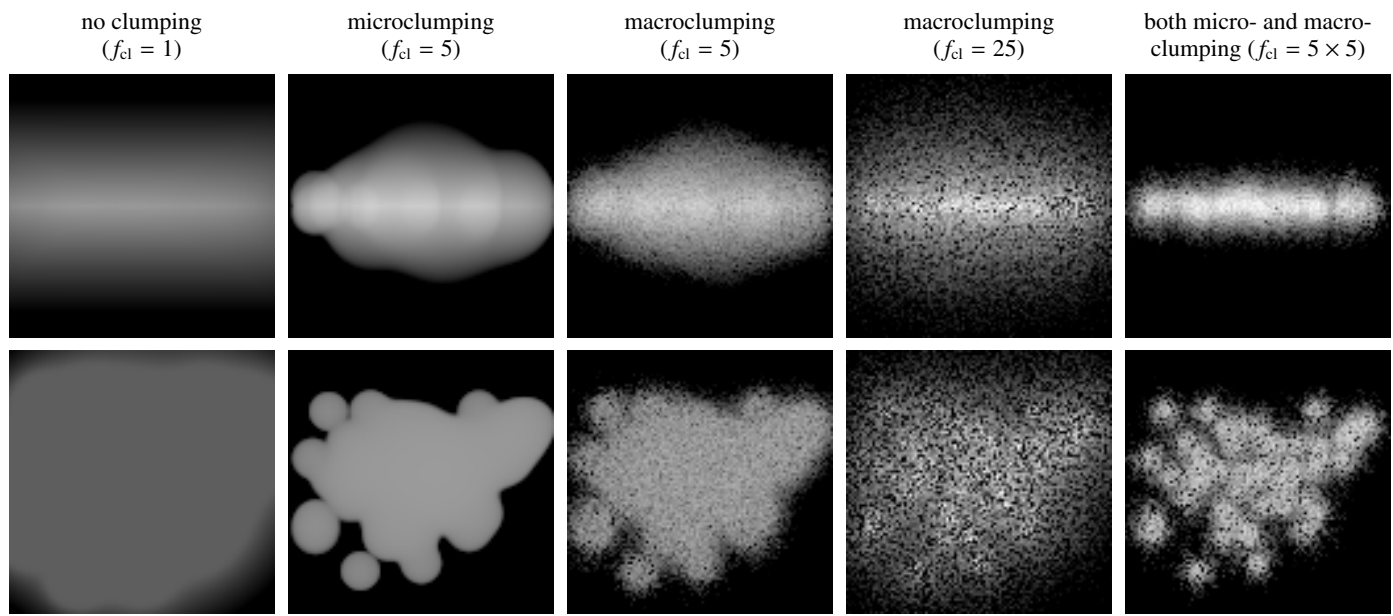

**Fig. 4.** Edge-on and face-on emission measure maps for the different simulated clumping scenarios (see text). In each case we have assumed a mean hydrogen number density of $n_{\rm H}(h) = 0.1\,{\rm cm}^{-3} \cdot e^{-|h|/h_0}$, where $|h|$ is the distance from the galactic plane and $h_0 = 1\,{\rm kpc}$ is the scale height of the gas. The simulations also share the same distribution and luminosities of the ionizing sources. The edge lengths of the maps are 5 kpc. The macroclumping case with $f_{\rm cl} = 5$ appears generally very similar to the microclumping case with the same clumping factor, as is expected for sufficiently fine grid resolution. For $f_{\rm cl} = 25$, the macroclumping becomes very porous at the used grid resolution, opening up channels devoid of gas that allow radiation to escape from the immediate environment of the sources without being absorbed. The combined micro- and macroclumping case with $f_{\rm cl} = 5 \times 5$ is very different from the macroclumping case with $f_{\rm cl} = 25$, giving an overall appearance resembling that of the cases with $f_{\rm cl} = 5$, but with the emission in each of the ionized bubbles more concentrated, somewhat analogous again to the transition from $f_{\rm cl} = 1$ to $f_{\rm cl} = 5$.

For the density of $10\,{\rm cm}^{-3}$ assumed in our spherically-symmetric computations in Sect. 2, the radii of the H II regions are on the order of the spatial grid cell size of $\Delta x \sim 50\,{\rm pc}$ used in the 3D simulations here,[19] and thus would not be resolved. For the remainder of this section, we concentrate on the diffuse gas and treat the H II regions around the ionizing sources as unresolved ("sub-grid") point sources that are not considered explicitly, but whose presence is implicitly accounted for by using the filtered stellar population SED as the ionizing flux for the DIG. Using our procedure of Sect. 2, we compute the filtered spectrum assuming a homogeneous H II region with $n_{\rm H} = 10\,{\rm cm}^{-3}$ that allows a fraction $f_{\rm esc} = 10\%$ of the ionizing flux to escape.

To study the effects of clumping and porosity on the ionization structure of the extraplanar gas, we perform a series of five simulations of the ionization of the DIG for a hydrogen number density that is similar to the values found in the diffuse ionized gas of the Milky Way. For all models, the mean density of the gas in the galactic plane is $n_{\rm H} = 0.1\,{\rm cm}^{-3}$ and its scale height is 1 kpc (Haffner et al. 2009 and references therein). After filter-ing by their H II regions, each of the sources provides an ionizing flux of $Q_{\rm H} = 7.1 \times 10^{49}\,{\rm s}^{-2}$ that can penetrate into the diffuse gas. As there are 50 ionizing sources in the simulated volume and the cross-section of that volume is $25\,{\rm pc}^2$, the total ionizing flux into the DIG is approximately $1.4 \times 10^{50}\,{\rm s}^{-1}\,{\rm kpc}^{-2}$. This value is similar to the $1.6 \times 10^{50}\,{\rm s}^{-1}\,{\rm kpc}^{-2}$ found by Barnes et al. (2015) to ionize the DIG in their model using a fractal density structure.

We note that this treatment is somewhat simplified, because, as shown in Sect. 2.3, the filtering of the ionizing radiation depends on the extent of the H II regions and the included stellar population. This means that the spectrum of the radiation that would escape if each of the stellar sources were individually surrounded by an H II region with a given escape fraction is not the same as the spectrum of the radiation escaping from a H II region surrounding a large association of stars. Compared to the energy distribution of the sources, the emergent spectrum is mainly modified by radiation hardening. Whereas for the sources themselves the number ratio between photons able to form He II and those able to form H II is $Q_{\rm He\,I}/Q_{\rm H\,I} = 1 : 6.7$, it rises to $1 : 3.75$ for the escaping radiation.[20] While the ratio in the unfiltered case is similar to the one found in 40 kK stars, the radiation hardening in the filtered case leads to a ratio similar to the one found in 45 kK stars. (A comparison of the ionizing fluxes of the used O star models is provided by Weber et al. 2015.) We show the spectral energy distribution of the filtered flux from the population synthesis model entering into the DIG in Fig. 5.

---
[19] In our simulation, the volume is subdivided into $101^3$ grid cells with a volume of $(49.5\,{\rm pc})^3$ each.

[20] A $Q_{\rm He\,I}/Q_{\rm H\,I}$ ratio of approximately 1:6 is required to fully ionize the helium within the hydrogen Strömgren sphere (Torres-Peimbert et al. 1974, Reynolds & Tufte 1995).

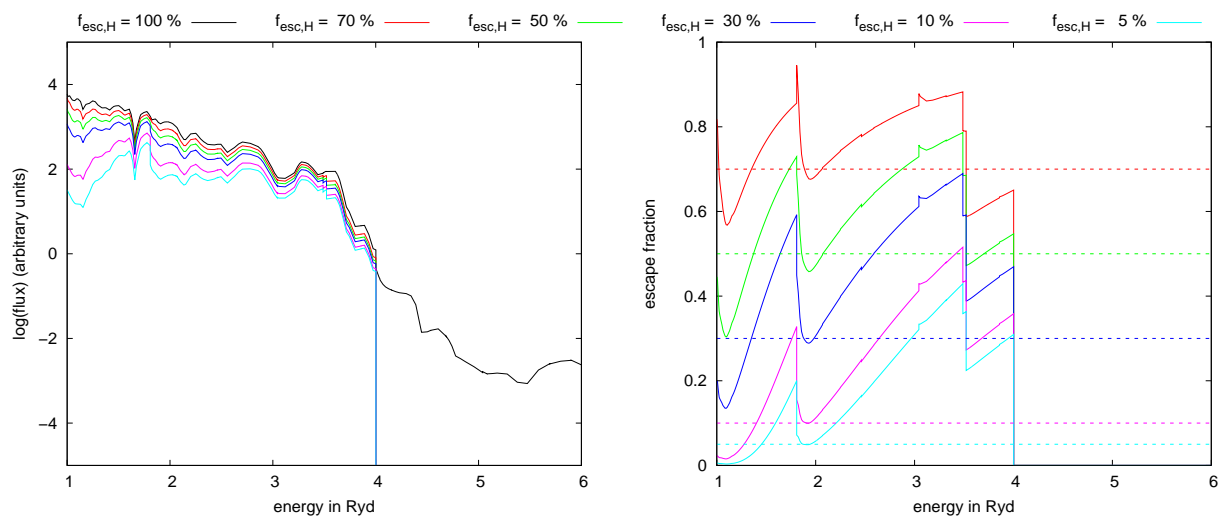

**Fig. 5.** *Left:* Spectral energy distributions from the population synthesis model, filtered as the stellar atmosphere models in Fig. 1. For our 3D simulations we use the spectral energy distribution of the model with $f_{esc} = 10\%$ (magenta) as ionizing flux for the DIG. *Right:* The corresponding energy-dependent escape fractions, as in Fig. 2. A comparison with Figs. 1 and 2 shows that the spectral energy distributions roughly resemble those using a 40 kK star.

For the density fluctuations in the diffuse gas we consider five different scenarios:

1. In the first simulation, the density structure of the gas is smooth, i.e., apart from the exponential drop of the hydrogen density for increasing heights above the galactic plane there are no inhomogeneities. (Note that the smooth density distribution refers only to the diffuse ionized gas itself; we still assume the same filtering of the ionizing radiation by gas in the environment of the sources as in the other simulations.)
2. In the second simulation, the gas has a clumping factor of $f_{cl} = \langle n_H^2 \rangle / \langle n_H \rangle^2 = 5$, a value similar to the one found in the observations by Reynolds (1991). We assume that 20% of the volume is filled with gas in which the hydrogen number density is $0.5\,\text{cm}^{-3}$, whereas the rest of the volume is empty. Here we use the "microclumping" approach, where the clumps are assumed to be optically thin and much smaller than the resolution elements of our simulation. The microclumping case is simulated numerically by multiplying the rate coefficients for the processes involving ion-electron interactions, namely recombination, collisional excitation, and thermal free-free radiation, by the clumping factor, as for otherwise constant conditions the electron number densities in the filled parts of the gas scale with the clumping factor.
3. In the third simulation, we replace the microclumping approach by the assumption of larger-scale inhomogeneities. This is realized by filling only 20% of the grid cells, selected randomly and independently of each other (so that in our simulations the typical length scale of the inhomogeneities corresponds roughly to the cell size of 50 pc), with gas with a density of $n_H = 0.5\,\text{cm}^{-3}$, leaving the rest of the cells empty. While we are certainly aware that this is not necessarily a realistic characterization of the actual statistical distribution of densities in a nonsmooth DIG, we have chosen this description here for two reasons. First, it is a natural extension of the unresolved (subgrid) microclumping model to a spatially resolved description, as both assume a two-phase medium consisting of swept-up matter and voids. Second, it is the most extreme clumping for a given spatial clumpiness scale. This and the smooth model represent the corner cases between which a more graduated description with intermediate density values would lie.
4. For comparison, we have also performed a simulation where we increased the (macro)clumping factor to $f_{cl} = 25$, i.e., only one out of 25 cells is filled with gas having a hydrogen number density of $n_H = 2.5\,\text{cm}^{-3}$ (in the galactic plane), while the rest of the cells remain empty.
5. Finally, we have simulated a combination of micro- and macroclumping, where both the total amount of gas within each of the grid cells corresponds to the macroclumping case with a clumping factor of $f_{cl} = 5$ and the internal structure of the gas within the single cells is characterized by an additional microclumping factor of 5.

We show both face-on and edge-on emission measure maps[21] of the simulation results in Fig. 4 and the ionization fractions of hydrogen as functions of the height above the galactic plane in Fig. 6. In the case of the homogeneous gas, ionized hydrogen exists within entire simulation volume, which reaches up to 2.5 kpc above the galactic plane. The ionization fraction actually rises at larger heights above the galactic plane. Due to the lower hydrogen densities, which lead both to lower recombination rates and to a lower opacity of the gas and therefore larger mean free paths of the ionizing photons.

The increased recombination rates in the denser clumps lead to a smaller ionized volume in both the microclumping and the macroclumping case with $f_{cl} = 5$. We find that in these cases the ionization fraction drops to approximately 10% (20% of the maximal value we obtain near the galactic plane) at a height of 1.2 kpc with only small differences between both cases. While clumping increases the recombination rates it also increases the fraction of the volume with below-average densities. Thus, chan-nels may form[22] that allow the ionizing radiation to reach larger distances from its source. This causes the gas to have an ionization fraction of approximately 25% even at large heights above the galactic plane for the $f_{cl} = 25$ case.

---
[21] The emission measure is defined as the integral $\int n_e^2\,ds$ of the square of the electron density along the line of sight. See also Weber et al. (2015) for details on the creation of these synthetic maps.

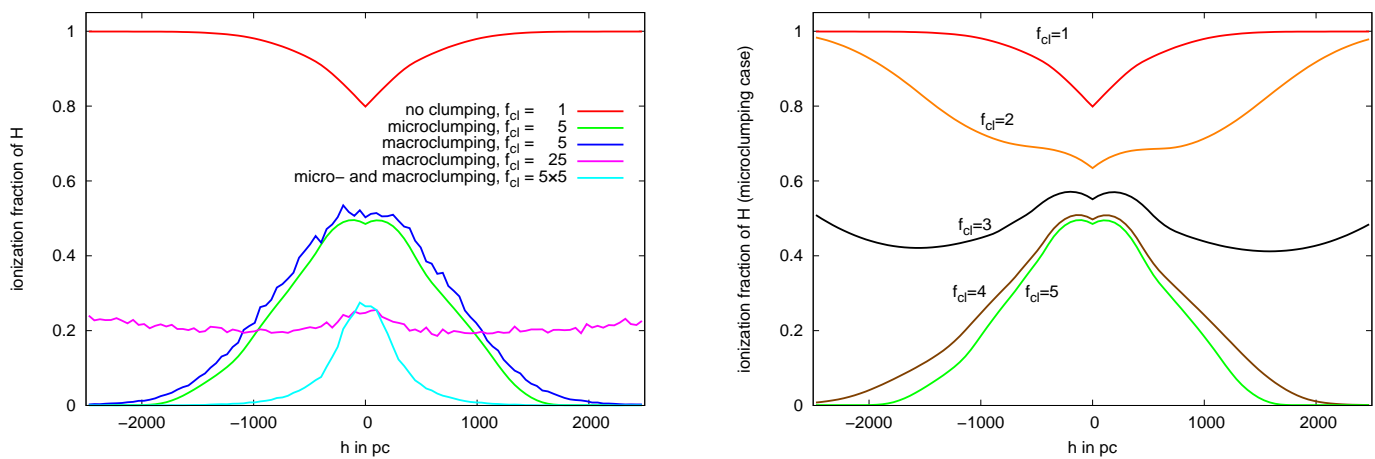

**Fig. 6.** *Left:* Ionization fractions of hydrogen as a function of the height above the galactic plane for the different clumping scenarios presented in Fig. 4. For the homogeneous density distribution, the gas is predominantly ionized throughout the simulation volume and the ionization fraction rises with increasing galactic heights due to the lower gas densities, which results in decreasing recombination rates and opacities. Both the microclumping case and the macroclumping case with $f_{cl} = 5$ are characterized by a similar dependence on the height above the galactic plane, reaching approximately 0.5 near the galactic plane and dropping to 0.1 at a height of about 1.2 kpc. In contrast, in the macroclumping scenario with $f_{cl} = 25$ the ionization fraction varies only slightly throughout the simulation volume. The combination of micro- and macroclumping shows the lowest ionization fraction of all the scenarios, due to the increased recombination rates in the filled cells. While the total clumping factor equals that of the $f_{cl} = 25$ macroclumping case, there are fewer extended empty regions through which radiation can penetrate to greater distances from the galactic plane, where less ionizing radiation is required to maintain the same fraction of ionized hydrogen due to the lower gas densities.
*Right:* As before, but for a parameter study varying the microclumping factor between 1 and 5. The ionization fraction rises at larger galactic heights in the case of low clumping, whereas it decreases and the gas eventually becomes neutral for high clumping factors. For immediate clumping factors, which for our geometry and density are around $f_{cl} \sim 3$, the ionization fraction as a function of height behaves in a non-monotonic way, first dropping to a minimum and then rising again.

The combined micro-and macroclumping case is characterized by the lowest ionization fractions among the simulations. The recombination rates in the filled parts of the volume are equal to the $f_{cl} = 25$ case described above, but due to the larger number of filled cells there are less channels for the radiation to leak out to large distances from the galactic plane.

Fig. 6 also visualizes the results of a further parameter study where we have assumed the microclumping scenario and varied the clumping factor between 1 (the unclumped case) and 5 (the microclumping case presented before). The mean densities and ionizing sources are the same as in the previous simulations. In the parameter study the ionization fraction of the gas as a function of the galactic height strongly depends on the clumping factor of the gas. For higher clumping factors, the ionization fraction decreases for larger galactic heights until the gas becomes neutral. In contrast, for low clumping factors the ionization fraction at large galactic heights is larger than within the galactic plane and the ionized region becomes matter-bounded, i.e., ionizing radiation can escape into the intergalactic space. For intermediate clumping factors, the ionization fraction drops until a minimum is reached and then rises again for larger distances from the central plane. The height-dependence of the ionization fraction results from several competing effects. On the one hand, photons have to penetrate the denser, more opaque, layers of gas to ionize high-altitude gas, so that the number of available ionizing photons decreases with increasing height. On the other hand, the recombination rate is lower in the thinner gas above the galactic plane. This increases the equilibrium value for the ionization fraction at higher altitudes. Furthermore, the opacity of the thin gas at large galactic height is lower than the opacity near the galactic plane. A comparison between the results for a clumping factor of 3 and for a clumping factor of 4 shows that the transition between clearly matter-bounded and clearly radiation-bounded ionized gas occurs within a relatively small range for the clumping factor. It can be expected that the ionization structure of the gas at large galactic heights is also strongly affected by slight variations of parameters of the gas that influence the recombination rates, such as the mean density or the temperature, or by a small variation of the amount of ionizing radiation. We note that the numerical values shown in the plots represent lower limits for the ionization fractions, as we assume open boundary conditions, thus neglecting photons entering the volume from the sides.

For the simulations with a clumping factor of $f_{cl} = 5$ and the simulation with combined micro- and macroclumping, we show line ratio maps of the diagnostically important optical lines [N II], [O II], and [O III] relative to H$\beta$ in the face-on view in Fig. 7 and in the edge-on view in Fig. 8. For the reasons discussed in Sect. 2, higher temperatures due to radiation hardening and a larger fraction of singly ionized nitrogen and oxygen, the [N II]/H$\beta$ and [O II]/H$\beta$ line ratios reach their maximal value in the outer regions of the ionized gas.

Barnes et al. (2014) show both the [N II]/H$\alpha$ ratios from their simulations as well as the corresponding observations of the DIG

---
[22] Note that this only happens if the spatial scale on which the medium is clumpy approaches the scale on which the medium becomes optically thick (i.e., the Strömgren radius) (cf. Wood et al. 2005, Owocki et al. 2004). As can be seen by the similarity of the microclumping and macroclumping models with $f_{cl} = 5$, and the subsequent dissimilarity to the $f_{cl} = 25$ macroclumping model, our grid resolution is sufficient to model the onset of this effect. Clumpiness on smaller scales can be adequately represented by microclumping.

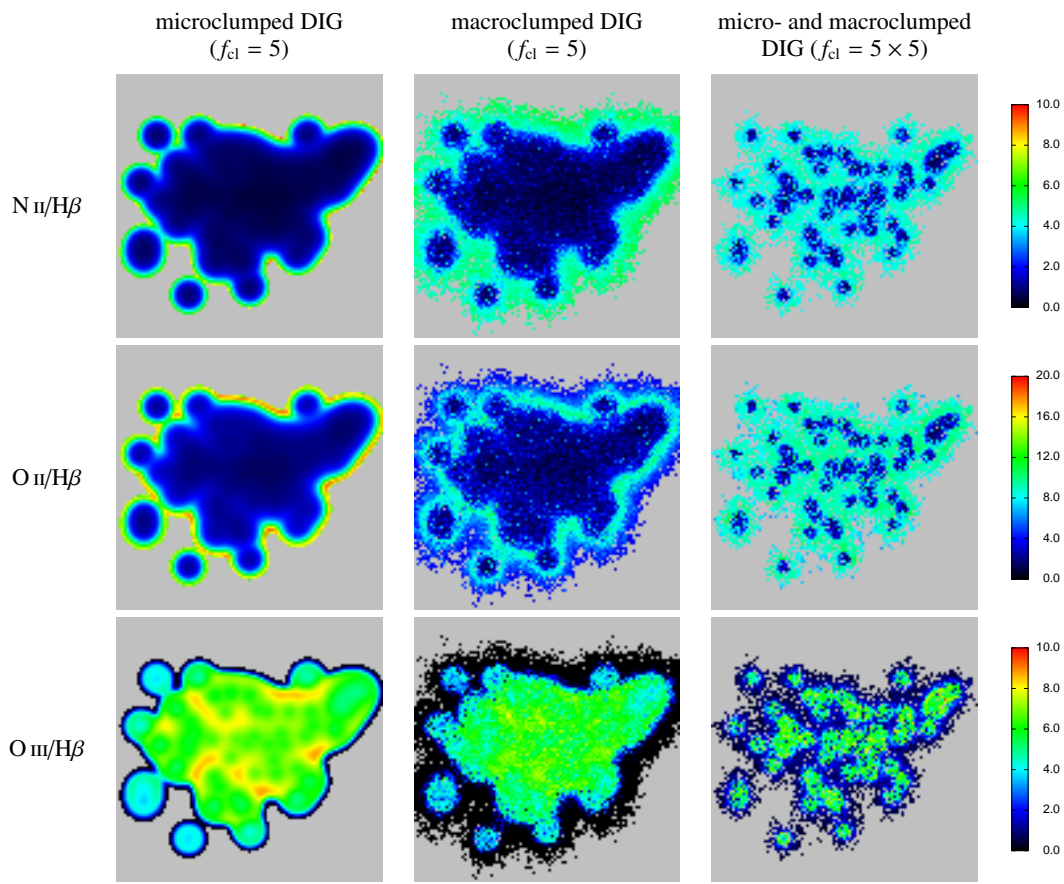

**Fig. 7.** Line ratio maps of several collisionally excited lines relative to H$\beta$ in the face-on view of the simulated spiral arm for different density structures of the gas. The gray color indicates emission measures less than 1 pc cm$^{-6}$.

around the Perseus arm and the inter-arm medium. For simulations that do not assume any heating process other than photoionization they obtain [N II]/H$\alpha$ ratios of approximately 0.4. The H$\alpha$/H$\beta$ ratio being approximately 2.9 without strong dependence on the gas density, this implies an [N II]/H$\beta$ ratio close to unity. Additionally, Barnes et al. have performed simulations including an additional heating term proportional to the electron density, $G_{\rm add} = n_{\rm e} \cdot 1.5 \times 10^{-27}$ erg s$^{-1}$. For these models they found almost unchanged emission line ratios for the relatively dense gas, but an increase of the [N II]/H$\alpha$ ratio to values close to or even above unity for more dilute gas.

Our 3D models do not directly correspond to the models of Barnes et al. (2014), both with respect to the ionizing radiation field and the density distribution of the gas. In particular, the Starburst99 model also considers hotter stars than the 35 000 K model star used by Barnes et al. to describe the radiation field, leading to a harder ionizing spectrum. As result, in our case the effect of the lower N II (and higher N III) abundances near the galactic plane is more pronounced, and it is not surprising that we obtain stronger variations of the [N II]/H$\beta$ as a function of the galactic height. We find a strong rise of the [N II]/H$\beta$ ratio near the edge of the ionized region, where in the microclumping case values of up to approximately 8 are reached, roughly corresponding to the maximal observed values in the environment of the Perseus arm, but clearly above the range of values found in the interarm medium. In the central parts of the ionized region, [N II]/H$\beta$ remains below two. For the macroclumping cases there is a less steep increase of the [N II]/H$\beta$ ratio towards the edge of the ionized volume.

We find a better agreement to the observational measurements of DIG shown in the same paper, especially regarding the rise of that ratio at larger galactic heights, but observations of the DIG of the Milky Way (Reynolds & Tufte 1995) suggest the absence of ionizing photons with energies beyond the ionization edge of neutral helium, which would also imply that there is very little doubly ionized nitrogen.

The [O III]/H$\beta$ ratio rises from the galactic plane to its maximal value at heights of up to 800 pc. For larger distances from the galactic plane the [O III]/H$\beta$ ratio decreases again. The outermost parts of the ionized gas do not emit [O III] line radiation. The overall O III/H$\beta$ ratios obtained by our 3D simulations are clearly too high to be consistent with the low abundance of ions requiring photons above the ionization threshold of neutral helium to be generated as found, for instance by Reynolds (1985) and Madsen et al. (2006). This could be explained by a lower escape fraction from the galactic plane and a significant contribution of other heat sources, like magnetic reconnection or shocks (cf. Sect. 1) or by higher escape fractions of ionizing radiation from H II regions around relatively cool ionizing sources (as discussed by Reynolds & Tufte 1995). Furthermore the contribution of the most massive and therefore hottest stars may be lower than assumed in our simulation. This could be either caused by a lower number of stars at the high-mass end of the mass function (for our simulations we assumed a Kroupa IMF with a mass cutoff of 100 $M_\odot$) or these massive stars may form in larger regions of overdense gas reducing the fraction of radiation being able to escape into the diffuse ionized gas.

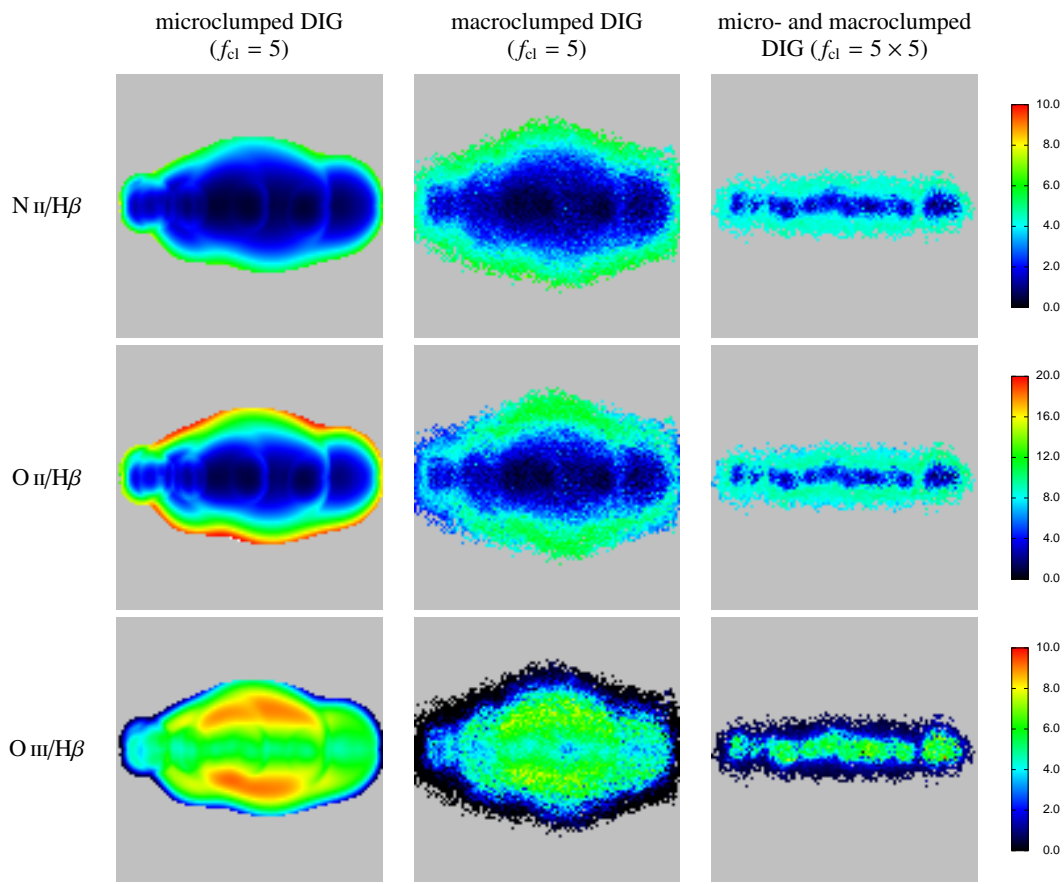

**Fig. 8.** As Fig. 7, but an edge-on view. Oxygen and nitrogen are predominantly doubly ionized close to the ionizing sources, and singly ionized only in the outer part of the ionized region. Thus, and because of radiation hardening and therefore higher temperatures, the maximum of the line ratios [N II]/H$\beta$ and [O II]/H$\beta$ occurs at larger heights above the galactic plane. For the cases with a clumping factor of $f_{\rm cl} = 5$ the [O III]/H$\beta$ ratio rises up to a height of approximately 700 pc above the galactic plane and drops again at larger distances.

In our simulations we also do not find an increase of the [O III]/H$\beta$ ratio at very large heights above the galactic plane as observed in edge-on galaxies by Otte et al. 2001 and Collins & Rand 2001, so that additional sources of ionization at larger distances from the galactic plane, such as central stars of planetary nebulae or white dwarfs as proposed by Sokolowski & Bland-Hawthorn (1991) and Flores-Fajardo et al. (2011) and supported by the observational results for galaxies with low star-formation by Lacerda et al. (2018) might need to be considered.

In the simulation with the homogeneous density (not shown in Figs. 7 and 8) the gas is almost fully ionized and optically thin, so that the effects of photons entering the volume from the sides can not be disregarded. Since we have used open boundary conditions in our simulations these photons are not considered, possibly leading to different results for the ionization structure of metals and consequently for the line emission. In the simulation with $f_{\rm cl} = 25$, on the other hand, most of the grid cells containing gas are optically thick for the ionizing radiation, so that the ionization structure and the properties of the radiation field would in reality vary significantly within the volume of a cell. Because a grid-based approach such as ours assumes that the temperature and ionization structure as well as the radiation field within each cell is constant, this could lead to inaccurate values for the computed emission line ratios.

### 3.2.1. A 3D simulation assuming a softer ionizing SED

To test the effects of a softer ionizing radiation on the line emission, we have repeated our simulation number 2 (the one using microclumping with $f_{\rm cl} = 5$) using as ionizing flux an unfiltered 35 000 K dwarf star SED. The reason for this choice is on the one hand a better comparability with the work of Barnes et al. (2014, 2015), and on the other hand earlier work by Reynolds & Tufte (1995) who discuss the possibility that the escape fraction tends to be higher for the radiation of cooler O (and even B) stars. We present the computed line ratio maps for this case in Fig. 9.

Comparing the results from this simulation with those using the filtered population SED, we find a considerably reduced [O III] emission, which is concentrated close to the galactic plane and disappears for large heights above the galactic plane due to the absence of doubly ionized oxygen. Observations that find a rising emission of [O III] for larger distances from the galactic plane can therefore clearly not be explained by a cool ionizing spectrum. By contrast, the emission of [N II] and [O II] rises slightly faster as function of the height above the galactic plane due to radiation hardening, the higher abundances of singly ionized nitrogen and oxygen, and the lower abundance of the very efficient coolant O III. However, the [N II]/H$\beta$ and [O II]/H$\beta$ ratios do not attain as large values because the softer ionizing spectrum leads to cooler gas temperatures, which in turn leads to increased recombination rates and a smaller ionized volume: while in the model using the hardened spectrum of the filtered population SED we find a mean temperature (weighted by square of the electron density) of approximately 9300 K, the corresponding weighted gas temperature in the model using the spectrum of the 35 000 K star is 7800 K.

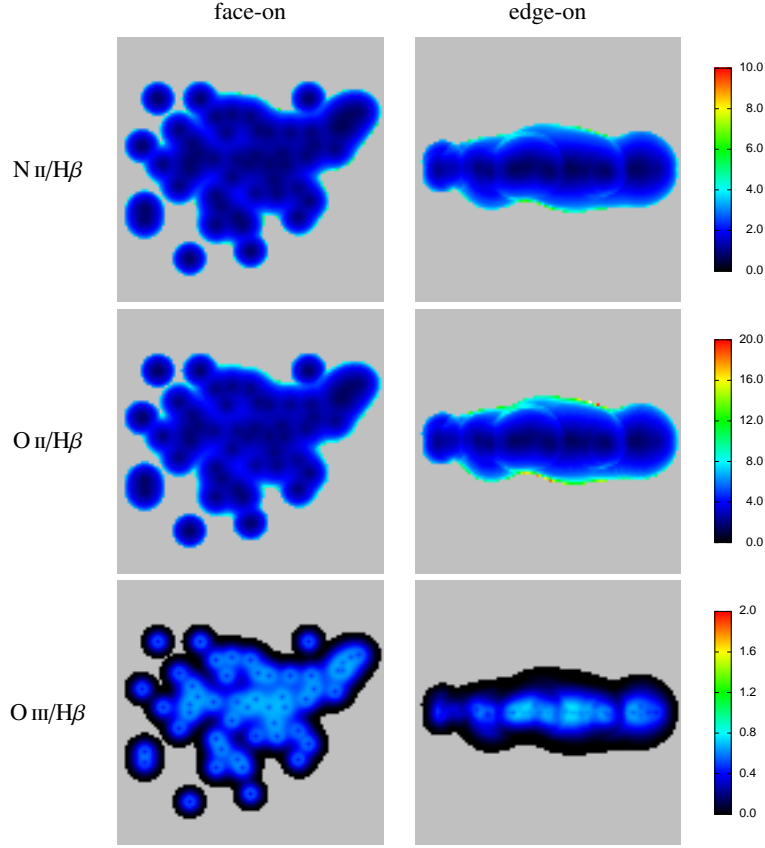

**Fig. 9.** Line ratio maps from microclumped models ($f_{cl}$ = 5) as in the leftmost columns of Figs. 7 and 8, using the same simulation volume and the same integrated hydrogen-ionizing flux as before, but taking the unfiltered spectrum of a 35 kK dwarf star with solar metallicity as the ionizing SED. The most notable difference to the models using the population SED is the weaker emission of the optical [O III] lines. The emission of the [N II] and [O II] lines is stronger close to the galactic plane, but shows a weaker rise close to the edge of the simulated volume.

### 3.3. Comparison with an analytical model for the ionization of the DIG

In this section we compute the height of the layer of ionized gas analytically and compare with the results from the numerical solutions presented above. Hereby we assume a microclumped gas whose mean density declines exponentially as a function of the height above the galactic plane and a purely plane-parallel geometry of the radiation field, without discrete sources, but with a homogeneous flow of ionizing photons emerging from the central plane. For simplicity, we further assume that the gas consists of pure hydrogen with a constant temperature (7500 K) in the ionized volume. For computing the ionization structure we use the Strömgren volume approximation that the gas is either fully neutral or fully ionized.

Under these assumptions we can analytically compute the height $z_{max}$ of the ionized layer of gas. In equilibrium, the emission rate $\dot{N}$ of ionizing photons equals the number of recombination processes per unit time. Normalized by surface area of the disk the equilibrium can be expressed as

$$\begin{aligned}
\frac{\dot{N}}{A} &= \int_{-z_{max}}^{z_{max}} \alpha_B n_{H\,II}(z) n_e(z) f_{cl}\, dz \\
&= \int_{-z_{max}}^{z_{max}} \alpha_B n_{H\,II}(0) e^{-|z|/h} n_e(0) e^{-|z|/h} f_{cl}\, dz \\
&= 2\alpha_B n_{H\,II}^2(0) f_{cl} \int_0^{z_{max}} e^{-2|z|/h}\, dz \\
&= \alpha_B n_{H\,II}^2(0) f_{cl} h \left(1 - e^{-2z_{max}/h}\right)
\end{aligned} \qquad (6)$$

where $\alpha_B$ is the (case-B) recombination coefficient, $n_{H\,II}$ the number density of ionized hydrogen and electrons, $f_{cl}$ the clumping factor, and $h$ the scale height of the gas. Equation 6 can now be solved for $z_{max}$:

$$z_{max} = -\frac{h}{2} \ln\left(1 - \frac{\dot{N}/A}{n_{H\,II}^2(0) f_{cl} h}\right) \qquad (7)$$

which becomes infinite for

$$\dot{N}/A \geq \alpha_B n_{H\,II}^2(0) f_{cl} h, \qquad (8)$$

in which case the ionized gas layer around the disk is no longer radiation-bounded and ionizing radiation can escape into intergalactic space.

For the parameters used in our numerical simulations, namely an ionizing flux of $\dot{N} = 3.55 \times 10^{51}$ s$^{-1}$, a section of the galactic plane of $(5\,\text{kpc})^2$, and a (mean) central hydrogen density of $0.1\,\text{cm}^{-3}$, the resulting height of the ionizing layer is $z_{\text{max}} = 221$ pc when assuming microclumping with $f_{\text{cl}} = 5$ and a recombination coefficient of $\alpha_B = 2.8 \times 10^{-13}\,\text{cm}^3\,\text{s}^{-1}$. The clumping factor below which the ionization becomes matter-bounded for otherwise unchanged conditions is $f_{\text{cl}} = 1.73$. A comparison of these values with our simulation results shows that for a purely plane-parallel geometry the maximal height of the ionizing gas is lower than in the numerical models that assume discrete sources (compare Fig. 6). The clustering of the ionizing sources in the spiral arm locally causes stronger ionizing fluxes of ionizing radiation than the plane-parallel model, allowing the ionizing radiation to reach larger heights above the galactic plane. Conversely, there are also regions in the galactic plane devoid of ionizing sources in the 3D models, so that the gas in these regions remains unionized due to lack of nearby ionizing sources and the absorption by the denser gas at low galactic heights (cf. Fig. 6), which is not he case in the simplified analytic model.

## 4. Summary and outlook

The diffuse ionized gas is the major contributor of warm ionized interstellar matter in disk galaxies, but the mechanisms that keep it ionized and determine its energy and density structure are not yet fully resolved. While a number of different ionization scenarios have been suggested, in this work we have concentrated on the ionization of the diffuse interstellar gas by radiation from massive hot stars. Due to their short lifetimes, these objects are associated with star formation regions and are thus concentrated near the galactic disk, so that their ability to ionize the diffuse gas depends on the ability of the emitted radiation to penetrate through the interstellar gas close to the disk.

When the radiation passes through the denser gas surrounding the sources, its attenuation is a function of the frequency-dependent absorption cross-sections of the atoms and ions contained within that gas. Thus the SED of the radiation that penetrates into the DIG can be significantly different from the SEDs of the stellar content of the traversed H II region. The modification of the SEDs depends on the original SED, the escape fraction, and the density distribution of the gas in the environment of the sources, most notably of hydrogen and helium, but also – to a lesser extent and depending on the metallicity of the gas – of carbon, nitrogen, and oxygen ions.

We have studied the dilution effects using both a sophisticated radiative transfer scheme that assumes the problem to be spherically symmetric, as well as a more simplified 3D-approach that, however, dispenses with the requirement of spherical symmetry. For the spherically symmetric case we have computed model grids of the line emission from the matter-bounded H II regions and from the diffuse gas ionized by radiation escaping from these H II regions. For these grids we have used SEDs from state-of-the-art stellar atmosphere models, similar to those which have already previously shown consistently good results as ionizing sources in numerical models of H II regions that were compared with observations. We vary the effective temperature ($30\,000\,\text{K} \leq T_{\text{eff}} \leq 50\,000\,\text{K}$) and the metallicity ($0.1\,Z_\odot \leq Z \leq 1.0\,Z_\odot$) of the model stars, as well as the escape-fraction ($0.05 \leq f_{\text{esc}} \leq 0.7$) of the ionizing radiation escaping from the H II regions around these stars.

The range of line ratios predicted by the grid of models encompasses the gamut of observed line ratios and thus supports the idea that the DIG is for the most part ionized by (filtered) radiation of hot stars in the galactic plane. As expected, we find that the attenuation accounts for two – partly opposing – effects that result in differences between the emission line spectra of the H II regions and of the diffuse ionized gas. The first effect is radiation hardening: due to the preferential absorption of photons near the ionization thresholds, the ionizing photons that escape from the H II regions are on average more energetic than the photons absorbed in the environment of the sources. Thus, photoionization heating leads to higher temperatures in the DIG than in H II regions, which in turn results in stronger emission of collisionally excited optical lines like N II, O II, O III, or S II.

Second, for stars with effective temperatures $T_{\text{eff}} \lesssim 35\,000$ K the volume in which helium is ionized is smaller than the corresponding volume of hydrogen, so that an H II region ionized by such stars (or by a group of stars whose SED is dominated by such stars) can be matter-bounded with respect to hydrogen, but radiation-bounded with respect to helium. In this case the escape fraction of helium-ionizing photons is significantly lower than the escape fraction of hydrogen-ionizing photons, so that the line emission from ions such as O III that require photon energies above the ionization threshold of He I to be formed is reduced or nonexistent in gas ionized by the filtered radiation.

This allows our model DIG to show similar properties as the diffuse ionized gas in the Milky Way, whose observed spectra indicate that both nitrogen and oxygen are predominantly singly ionized and helium appears to be mostly neutral. Nevertheless, stellar radiation with a low number of He-ionizing photons as the sole source of ionization would not be able to account for an increasing [O III]/H$\beta$ ratio at large distances from the galactic plane as has been observed in the DIG of some edge-on galaxies.

The DIG itself also acts as a filter for the regions farther away from the ionizing sources. To study the effects of different distributions of inhomogeneities in the DIG, we have performed a series of 3D simulations representing a section of the galactic disk around a spiral arm and the diffuse gas above and below it. The models share a common exponential density distribution of the gas but differ in the degree and type of inhomogeneities. We consider three different scenarios: homogeneous gas, inhomogeneous gas with "microclumping" (i.e, the inhomogeneities are optically thin and not spatially resolved within the simulation), and inhomogeneous gas with macroclumping, where we have chosen the inhomogeneities to be of similar scale as the resolution element of the simulation of approximately 50 pc. For the ionizing sources we have used an SED computed using the stellar population synthesis code Starburst99 corresponding to the star formation rate in a spiral arm and a Kroupa IMF. In our simulation, the ionizing flux is equally distributed among a set of sources that are positioned so that their spatial distribution roughly corresponds to those in a spiral arm, and additionally modified by homogeneous H II regions with an escape fraction of 0.1.

We find that small-scale inhomogeneities, as in microclumping, or in macroclumping where the average distance between clumps is smaller than the mean free path of photons in a homogeneous medium, considerably reduce the ability of ionizing photons to escape from the environment of the galactic disk. However, macroclumping with high clumping factors leads to the formation of channels through which the radiation is able escape from the environment of the galactic disk without being absorbed.

We present synthetic line ratio maps that show the run of the line ratios as function of galactic height as well as illustrate the spatial variability of the line ratios resulting from the non-regular distribution of the ionizing sources (which obviously depends on their chosen arrangement.) We find that the [O II]/H$\beta$ and [N II]/H$\beta$ line ratios increase for larger heights above the galactic plane. However, we find considerably larger [O III]/H$\beta$ ratios than observed in the Milky Way or in M 31, which may indicate that although the overall number of photons emitted by the stellar population in a star-forming galaxy would be sufficient to ionize the DIG, the filtering characteristics by the stellar environment of the hot massive stars may be different from those we assumed in this simulation.

For comparison purposes, and because previous results from other work as well as our spherically symmetric results indicate ionization by a softer radiation field, we have additionally performed a 3D simulation where we have changed the SED of the ionizing radiation to correspond to the unfiltered spectrum of a 35 kK model star. As expected from our spherically symmetric results, we have found a considerably better agreement with the observed line ratios.

To broaden the scope of simulations of the diffuse ionized gas, future work may combine 3D simulations with parameter studies regarding the properties of the stellar populations (metallicity, initial mass function, time-dependent effects like starbursts) and consider possible additional energy sources such as magnetic reconnection or hot evolved stars outside the galactic disk. It may also account for the distribution of the ionizing sources in individual galaxies.

*Acknowledgements.* We would like to thank R. Walterbos and the Astrophysical Journal for granting us permission to reproduce the observational images used in Fig. 3. Furthermore we would like to thank our referees for their useful suggestions that helped to improve this work. This work was supported by the *Deutsche Forschungsgemeinschaft (DFG)* under grants PA 477/18-1 and PA 477/19-1.